\documentclass{ptephy_v1}

\preprintnumber{XXXX-XXXX} 
\usepackage{amsthm}
\usepackage{bm,latexsym,amsmath,amssymb,amsfonts,bbold}
\usepackage{graphics}
\usepackage{ulem}

\begin{document}

\title{Fermionic Casimir effect in the presence of compact dimension in field theory with Lorentz invariance violation} 

\author{Ar Rohim$^{1}$ and Apriadi Salim Adam$^2$ }
\affil
{
$^{1}$Departemen Fisika, FMIPA, Universitas Indonesia, Depok 16424, Indonesia\\
$^{2}$Research Center for Quantum Physics, National Research and Innovation Agency (BRIN), South Tangerang 15314, Indonesia
}

\begin{abstract}
In this study, we investigate the effect of the Lorentz invariance violation on the Casimir energy and pressure of the massive fermion field in the presence of the compact dimensions with topological $R^4\times S^1$, referring to the Kaluza-Klein model.
In the system, the Dirac field is confined between two parallel plates with the geometry described by MIT bag boundary conditions, and the compactified dimension satisfies quasi-periodic boundary conditions. We investigate two directions of the Lorentz violation, namely, space- and time-like. The results reveal that in the space-like vector case, the Lorentz violation's strength and the extra dimension affect the Casimir energy and pressure. In contrast, in the time-like vector case, they are only affected by the extra dimension. We also propose an indirect method to estimate the size of the extra dimension by comparing the frequency shift of the massless fermionic case to that of the scaled experimental data for the electromagnetic field.
\end{abstract}

\subjectindex{A64, B30, B69}

\maketitle

\section{Introduction}
Several quantum gravity theories have been proposed to address the unification of quantum mechanics and gravity, such as string theory \cite{Kaku:1974zz, Kaku:1974xu} and loop quantum gravity \cite{Rovelli:1991zi, Rovelli:1997yv}. The proposed theories imply the Lorentz invariant violation at the Planck scale. Kaluza-Klein's theory, which proposes the unification of the electromagnetic (EM) field and gravity, leads to the existence of an extra dimension. In such a theory, the extra dimension is compactified and the Lorentz symmetry is violated (see Ref.~\cite{Rizzo:2005um}). In Ref. \cite{Carroll:2008pk}, it has been stated that the size of the extra dimension remains small (at Planck scale) by introducing the Lorentz-violation tensor field called the aether field. In such a scenario, the standard dispersion relation is modified. 
On top of that, the Lorentz invariance violation can appear in other mechanisms such as standard model extension \cite{Colladay:1998fq} and Ho${\rm \check{r}}$ava-Lifshitz theory \cite{Horava:2009uw}. The field theory involving Lorentz violation may offer new aspects in the phenomena of the Casimir effect, which has been experimentally confirmed with high precision.

The Casimir effect, which is a result of the quantum vacuum fluctuation, was first predicted in 1948 \cite{Casimir:1948dh}. It involves an EM field confined between two plates placed at very close distances, generating an attractive force. The effect was experimentally confirmed by Sparnay in 1958 \cite{Sparnaay:1958wg}, and subsequent experiments have achieved high precision in confirming this phenomenon \cite{Lamoreaux:1996wh, Mohideen:1998iz, Roy:1999dx, Bressi:2002fr}. The study of the Casimir effect has also been explored from a theoretical perspective for various fields, including scalar and fermionic fields, as well as the involvement of the magnetic field background \cite{Sitenko:2014kza, Cougo-Pinto:1998jwo, Ostrowski:2005rm, Elizalde:2002kb, Cougo-Pinto:1998jun, Erdas:2013jga, Farina:2006hk}. Besides that, the behavior of the Casimir effect is determined by the boundary conditions. For the type of the fermion field, boundary conditions with the MIT bag model \cite{Chodosetal1974a, Chodosetal1974b, Johnson1975} and its extension (see e.g. \cite{Lutken:1983hm} for the chiral MIT boundary conditions), can be applied to represent the confinement of the fields. As for the scalar field, one could use variants of the Dirichlet, Neumann  and/ or mixed boundary conditions \cite{Ambjorn:1981xw}.

The Casimir effect in the presence of the extra dimensions has been extensively studied in the literature (see e.g., Refs.~\cite{Perivolaropoulos:2008pg, Elizalde:2009nt, Bellucci:2009hh, Erdas:2021fwe, Kirsten:2009kx, Teo:2009tm, Poppenhaeger:2003es} and the references therein). Further, a study with thermal corrections for this system has also been discussed \cite{Khoo:2011ux}. Most studies have been conducted within the framework of standard field theory, preserving Lorentz symmetry, except for the discussion in Ref. \cite{deFarias:2023xjf}, which addresses Lorentz invariance violation in the EM field. To our knowledge, no such investigation has been carried out for the Dirac field. Therefore, in this study, we employ the model developed in Ref. \cite{Bellucci:2009hh} but work within the framework of quantum field theory with Lorentz symmetry violation, focusing on the aether field \cite{Cruz:2018thz, Carroll:2008pk, Gomes:2009ch, deFarias:2023xjf} (see Refs.~\cite{Ferrari:2010dj, MoralesUlion:2015tve, daSilva:2019iwn, deMello:2024lmi} for the study on the Casimir effect with Lorentz invariant violation in Ho${\rm \check{r}}$ava-Lifshitz theory).

The Dirac field type with Lorentz symmetry breaking has also attracted some attention from the condensed matter field, especially the experimental side. A few years ago, Xu et al. revealed a specific type of particle called Type II Weyl fermions that violate the Lorentz symmetry principle \cite{doi:10.1126/sciadv.1603266}. Their study focuses on the material LaAlGe and suggests it might be the first discovery of these type II Weyl fermions. On the other hand, there was also an attempt to find type II Dirac fermion in a material PtTe2 \cite{Yan2017}. They claimed that this is the first experimental discovery of type-II Dirac fermions in a single crystal of PtTe2.  They utilized special techniques to study the electronic structure of the material and confirmed the presence of these unique particles. The discovery of type-II Dirac/Weyl fermions would be significant because it would challenge our current understanding of physics, particularly condensed matter physics and quantum field theory. 
These materials may exhibit exotic properties and lead to new discoveries about quantum phenomena and topological phases of matter \cite{PhysRevLett.108.140405,PhysRevB.89.235127}.

Based on the aforementioned motivations, in this work, we study a system of the Dirac field which is confined between two parallel plates in five-dimensional flat spacetime. We use the topology $R^4\times S^1$ that refers to the Kaluza-Klein model, where the extra dimension is compactified in a circle. In such a system,  we impose a condition in which the field must satisfy two kinds of boundary conditions, namely, MIT bag  \cite{Chodosetal1974a, Chodosetal1974b, Johnson1975} and quasiperiodic boundary conditions. We aim to investigate the roles of the Lorentz violation and the presence of extra dimensions on the Casimir energy and its pressure. We consider two Lorentz violation directions, i.e., time- and space-like directions. In particular, the space-like directions correspond to the uncompactified and compactified components. 

We also probe the size of the extra dimension. For an EM field, one can obtain the constraint by comparing theoretical results with the experimental data \cite{deFarias:2023xjf}. Unfortunately, the experimental data for the Casimir effect of the fermionic field have not been provided yet. In this study, we compare the theoretical result of the massless fermionic field in the standard 1+3-Minkowski space time to that of an EM field as the experimental reference for the Casimir effect of the fermionic field. This approach allows us to investigate the roles of the Lorentz violation and the extra dimension.

The rest structure of this paper is organized as follows. In Section~\ref{model}, the description of our model is presented. In Section~\ref{CasimirEnergy}, we derive the expression of the Casimir energy for the space-like vector cases, whereas, for the time-like vector case, it can be obtained by taking particular conditions of the space-like one. The derivation of the Casimir pressure is given in Section~\ref{CasimirPressure}. In Section~\ref{ExtraDim}, the size of the extra dimensions obtained from taking the first derivative of the Casimir pressure with respect to the plate's distance is examined. Section~\ref{Summary} is devoted to our summary and outlook. In the present study, we use the natural units with $\hbar=c=1$, unless necessary.
\section{Model}
\label{model}

We consider the Dirac field confined between two parallel plates in the five-dimension flat spacetime background with coordinates $x^a=(x^\mu,x^5)$ and $\mu=0,1,2,3$. We discuss such a system in the case of the presence of compact dimensions with topological $R^{4}\times S^1$, where $R^4$ corresponds to the 1+3 usual Minkowski spacetime and $S^1$ corresponds to the extra dimension which is compactified on a circle. The line element of the spacetime background is given by the following: 
\begin{eqnarray}
ds^2=g_{\mu\nu}dx^\mu dx^\nu- (dx^5)^2,
\label{lineelement}
\end{eqnarray}
with $g_{\mu\nu}(\equiv {\rm diag.}(+1,-1,-1,-1))$ is the usual  metric tensor for $1+3$ dimensional Minkowski spacetime. The Lagrangian density for the Dirac field $\Psi$ with mass $m$ involving Lorentz invariant  violation is given by the following\cite{Cruz:2018thz}:
\begin{eqnarray}\label{lagden}
{\cal L}=\bar \Psi(i\gamma^a \partial_a-m+i\lambda u^a u^b\gamma_a \partial_b)\Psi,
\end{eqnarray}
where $\gamma^a(\equiv(\gamma^\mu, \gamma^5))$ are $4\times 4$ gamma matrices, $\lambda$ is a dimensionless parameter that determines how large the violation is, 
and $u^a$ 
is a constant five vector that determines the direction of the Lorentz violations. 
In the present stdudy, we use gamma matrices as follows \cite{Sanchez:2011qb}
\begin{eqnarray}
 \gamma^0=  \begin{pmatrix}
{\mathbb 1} & 0\\
0 &-{\mathbb 1}
\end{pmatrix},
~~~~
 \gamma^j= \begin{pmatrix} 0 & \sigma^j\\
-\sigma^j &0
\end{pmatrix},~~{j=1,2,3}, 
~~{\rm and}~~
\gamma^5= \gamma^0\gamma^1\gamma^2\gamma^3,
\end{eqnarray}
where $\mathbb{1}$ represents the $2\times 2$ identity matrix, the matrices $\sigma^j$ are the $2\times 2$ Pauli matrices satisfing the anti-commutation relations as $\{\sigma^j,\sigma^k\}=2\delta_{jk}$ and the gamma matrices $\gamma^\mu$ satisfy the anti-commutation relation as $\{\gamma^\mu,\gamma^\nu\}=2g^{\mu \nu}$. Here, gamma matrix $\gamma^5$ anti-commutes with each gamma matrices $\gamma^\mu$, i.e., $\{\gamma^5,\gamma^\mu\}=0$.

The field  $\Psi$  satisfies the modified Dirac equation as follows:
\begin{eqnarray}
[i\gamma^a \partial_a-m+i\lambda u^a u^b\gamma_a \partial_b]\Psi=0. 
\end{eqnarray}
 The positive- and negative-frequency solutions that satisfy the above Dirac equation are given by the following:
\begin{eqnarray}
\psi^{(+)}_\beta&=&{\cal N}_\beta e^{-i\omega t} 
\begin{pmatrix}
\chi^{(+)}_1\\
\chi^{(+)}_2
\end{pmatrix},\\
\psi^{(-)}_\beta&=&{\cal N}_\beta e^{i\omega t}
\begin{pmatrix}
\chi^{(-)}_1\\
\chi^{(-)}_2
\end{pmatrix}, 
\end{eqnarray}
respectively, where ${\cal N}_\beta$ is the normalization constant which can be determined by applying the orthonormality conditions and $\chi_{1(2)}$ corresponds to the upper(lower) two-component spinors. 

In the system, the  compact dimensions satisfy the quasiperiodic boundary conditions given as follows \cite{Bellucci:2009hh}:
\begin{eqnarray}
\Psi(x^\mu,x^5+q)=e^{2\pi i\beta}\Psi(x^\mu,x^5),
\label{BCcompactified}
\end{eqnarray}
where $q$ is the length of the compactification of the extra dimensions.  The parameter $\beta$ corresponds to the value of the phase space.  It  has  an interval value of $0\leq\beta<1$, where $\beta=0$ and $\beta=1/2$ corresponds to periodic and antiperiodic boundary conditions, respectively \cite{deFarias:2023xjf}. The Dirac field is confined between two parallel plates placed at $x^3=0$ and $x^3=a$. The properties of plates are explained by the boundary conditions of the MIT bag model given as follows \cite{Chodosetal1974a, Chodosetal1974b, Johnson1975}:
\begin{eqnarray}
in_{a}\gamma^{a}\Psi=\Psi,
\label{MITBagBC}
\end{eqnarray}
where $n_a$ is the normal unit vector perpendicular to the boundary surface. The above MIT bag boundary conditions give property in which the normal currents  
must vanish at the boundary surfaces of the plates. Imposing the field to the above boundary conditions leads to the discretized perpendicular momentum, as will be shown below.

\section{Casimir energy}
\label{CasimirEnergy} 
In this section, the Casimir energy for two vector cases, namely, time- and space-like vectors, will be discussed following the study in Ref.~\cite{Cruz:2018thz}. For the space-like cases, we consider two directions of Lorentz violation, namely, $x^3$- and $x^5$-directions related to the uncompactified and compactified components, respectively. We will show that the Lorentz violation in the time-like case does not affect the Casimir energy and its pressure, as pointed out in Refs.~\cite{Cruz:2018thz, Rohim:2023tmy}. By contrast, for the space-like vector cases, the Lorentz violation affects both Casimir energy and its pressure. We investigate the solution for the Dirac field that satisfies the Dirac equation under MIT bag model \cite{Chodosetal1974a, Chodosetal1974b, Johnson1975} and quasi-periodic boundary conditions. We will show that the corresponding momentum $k_3$ and $k_5$ with their directions will be discretized due to the presence of the boundary conditions, whereas the other momentum components remain continuum. We derive the expression of the vacuum energy of the Dirac field for each vector case. To obtain the expression of the Casimir energy, we apply the Abel-Plana like summation \cite{Romeo:2000wt} in the vacuum energy.

\subsection{Time-like vector case} 

In this subsection, we consider the time-like vector case $u^{(0)}=(1,0,0,0,0)$. In this case, we have the positive- and negative-frequency solutions as follows
\begin{eqnarray}
\psi^{(t,+)}_\beta&=&{\cal N}_\beta e^{-i\omega t} 
\begin{pmatrix}
\chi^{(t,+)}_1\\
{(-i\sigma^j\partial_j+\partial_5)\chi^{(t,+)}_1 / ((1+\lambda)\omega+m)}
\end{pmatrix},\\
\psi^{(t,-)}_\beta&=&{\cal N}_\beta e^{i\omega t}
\begin{pmatrix}
{(i\sigma^j\partial_j-\partial_5) \chi^{(t,-)}_2/ ((1+\lambda)\omega+m)}\\
\chi^{(t,-)}_2
\end{pmatrix}, 
\end{eqnarray}
respectively, where the two component spinors are given by the following:
\begin{eqnarray}
\chi^{(t,+)}_1&=&e^{i{\bf k}_\parallel \cdot {\bf x}_\parallel} (\phi^{(t)}_{+}e^{ik_{3}x^{3}}+\phi^{(t)}_{-}e^{-ik_{3}x^{3}}),\\
\chi^{(t,-)}_2&=&e^{-i{\bf k}_\parallel \cdot {\bf x}_\parallel} (\varphi^{(t)}_{+}e^{ik_{3}x^{3}}+\varphi^{(t)}_{-}e^{-ik_{3}x^{3}}),
\end{eqnarray} 
with ${\bf k}_\parallel=(k_1,k_2,k_5)$ and ${\bf x}_\parallel=(x^1,x^2,x^5)$.
In this time-like vector case, we also note that the eigenfrequencies are given by the following:
\begin{eqnarray}
\omega=(1+\lambda)^{-1}\sqrt{m^2+k_1^2+k_2^2+k^2_3+k^2_5}.
\end{eqnarray}

In the absence of boundaries, all momenta are continuum, as mentioned above. However, in our system, we have MIT bag and quasiperiodic boundary conditions so that we have to analyze the behavior of the momenta under boundary conditions. At the first plate $x^{3}=0$, the normal surface is given by $n_a=(0,0,0,1,0)$. At this surface, by imposing the MIT bag boundary conditions in Eq.~\eqref{MITBagBC}, the following equations are obtained:
\begin{eqnarray}
\phi^{(t)}_+ &=& -{m((1+\lambda)\omega+m)+k^2_{3}-\sigma^3k_3(ik_5+\sigma^1k_1+\sigma^2k_2)\over (m-ik_{3})((1+\lambda)\omega+m)}\phi^{(t)}_-,
\label{phit1}\\
\varphi^{(t)}_-&=& -{m((1+\lambda)\omega+m)+k^2_{3}-\sigma^3k_3(-ik_5+\sigma^1k_1+\sigma^2k_2)\over (m+ik_{3})((1+\lambda)\omega+m)}\varphi^{(t)}_+.
\label{varphit1}
\end{eqnarray}
Imposing the same boundary condition at the surface of the second plate $x^{3}=a$ with the normal surface given by $n_a=(0,0,0,-1,0)$, we have
\begin{eqnarray}
\phi^{(t)}_+ &=& -{m((1+\lambda)\omega+m)+k^2_{3}-\sigma^3k_3(i k_5+\sigma^1k_1+\sigma^2k_2)\over (m+ik_{3})((1+\lambda)\omega+m)}e^{-2k_3a}\phi^{(t)}_-,
\label{phit2}\\
\varphi^{(t)}_-&=& -{m((1+\lambda)\omega+m)+k^2_{3}-\sigma^3k_3(-ik_5+\sigma^1k_1+\sigma^2k_2)\over (m-ik_{3})((1+\lambda)\omega+m)}e^{-2k_3a}\varphi^{(t)}_+.
\label{varphit2}
\end{eqnarray}
Equalling $\phi^{(t)}_\pm$ in Eqs.~\eqref{phit1} and \eqref{phit2}, we will arrive at the condition where the momentum $k_{3}$ must satisfy the following constraint
\begin{eqnarray}
ma \sin(k_{3} a)+ k_{3} a \cos(k_{3}a)=0.
\label{ConstraintMomentum}
\end{eqnarray}
The same constraint will be obtained when one comparing $\varphi^{(t)}_\pm$ in Eqs.~\eqref{varphit1} and \eqref{varphit2}, which means that both positive- and negative-frequencies solutions have the same $k_3$.  The solution of the above constraint is $k_{n}(\equiv k_3 a)$ with $n=1,2,\cdots$. One can see that the Lorentz violation does not affect the momentum $k_n$, as has been revealed in Ref.~\cite{Cruz:2018thz}.  In the massless case, the constraint becomes $\cos(k_3 a)=0$, so the solution is given by $k_n=(n-1/2)\pi$.  Imposing the quasiperiodic boundary conditions \eqref{BCcompactified}, we have the discretized momentum $k_5$ as follows
\begin{eqnarray}
k_{\ell}={2\pi \over q} (\ell+\beta),~~\ell=0,\pm 1, \pm 2, \cdots.
\label{k5discrete}
\end{eqnarray}
Then, due to the above  boundary conditions, the eigenfrequencies are also discretized as follows 
\begin{eqnarray}
\omega_{n,\ell}=\sqrt{m^2+k^2_1+k^2_2+\left({k_{n}\over a}\right)^2+k^2_{\ell}}.
\end{eqnarray}

The field expansion is given by
\begin{eqnarray}
\Psi^{(t)}=\int dk_1\int dk_2\sum_{n=1}^\infty \sum_{\ell=-\infty}^\infty \sum_{s=1}^2\bigg[ b_{k_1,k_2,n,\ell, s} \psi^{(+,t)}_{k_1,k_2, n,\ell, s} +d^\dagger_{k_1,k_2,n,\ell, s}  \psi^{(-,t)}_{k_1,k_2, n,\ell, s} \bigg], 
\label{fieldexptime}
\end{eqnarray}
where $s$ is related to the spin, and the creation and annihilation operators satisfy the anti-commutation relations as follows
\begin{eqnarray}
\lbrace b_{k_1,k_2,n,\ell, s}, b^\dagger_{k'_1,k'_2,n',\ell', s'}\rbrace=\lbrace d_{k_1,k_2,n, \ell, s}, d^\dagger_{k'_1,k'_2,n',\ell', s'}\rbrace=\delta (k_1-k'_1)\delta(k_2-k'_2)\delta_{nn'}\delta_{\ell\ell'}\delta_{ss'},
\end{eqnarray}
and the other form of the anti-commutation relations vanishes. 
The positive- and negative-frequency solutions satisfy the orthonormality conditions as follows 
\begin{eqnarray}
\int d {\bf x}_\parallel \int_0^a dx^3 \psi^{(\pm,t)\dagger}_{k_1,k_2, n,\ell, s}  \psi^{(\pm,t)}_{k'_1,k'_2, n',\ell', s'} =\delta (k_1-k'_1)\delta(k_2-k'_2)\delta_{nn'}\delta_{\ell\ell'}\delta_{ss'}.
\label{othonormalitytime}
\end{eqnarray}

We next discuss the vacuum energy. From the Lagrangian density in Eq.\eqref{lagden}, we have the Hamiltonian in the case of time-like direction as follows,
\begin{eqnarray}
\hat H=\int d {\bf x}_\parallel \int_0^a dx^3 \bar \Psi^{(t)}(-i\gamma^j\partial_j+m)\Psi^{(t)}= (1+\lambda) i \int d {\bf x}_\parallel \int_0^a dx^3 \Psi^{(t)\dagger}\partial_t\Psi^{(t)}. 
\end{eqnarray}
Substituting the field expansion \eqref{fieldexptime}, the above Hamiltonian becomes
\begin{eqnarray}
\hat H= (1+\lambda) \int d k_1 \int d k_2  \sum_{n=1}^\infty  \sum_{\ell=-\infty}^\infty  \sum_{s=1}^2\omega_{n}\bigg[ b^\dagger_{k_1,k_2,n,\ell, s}b_{k_1,k_2,n,\ell, s} + d^\dagger_{k_1,k_2,n,\ell, s}d_{k_1,k_2,n,\ell, s}-\delta^{(2)}(0)\bigg]. 
 \label{hamilt}
\end{eqnarray}
where we have used the orthonormality conditions and the anti-commutation relations.
Then, the vacuum energy in the presence of boundary conditions and extra dimension with Lorentz violation in a time-like direction can be evaluated as  follows
\begin{eqnarray}
E_{\rm Vac.}=\langle 0|\hat H|0\rangle
=-{L^2\over 2\pi^2}\int {d k_1}\int {d k_2}\sum_{\ell=-\infty}^\infty\sum^\infty_{n=1}\sqrt{m^2+k^2_1+k^2_2+\left({k_{n}\over a}\right)^2+k^2_{\ell}}, \label{time-like-vac}
\end{eqnarray}
where we have used $L^2$ to represents the plate's area.
When taking the matrix element of the Hamiltonian with the vacuum state, we note that the main contribution only comes from the third term inside the bracket of Eq.~\eqref{hamilt} while the other terms vanish. Besides that, the Lorentz violation does not contribute to the vacuum energy, as has been explored in the earlier study by Refs.~\cite{Cruz:2018thz,Rohim:2023tmy}, in which they discussed the case in the absence of extra dimension. 
Therefore, as per the purpose of this study, which is to determine the effect of Lorentz violations and the extra dimension on the Casimir energy and pressure, the discussion on the time-like vector case ends here.
Moreover, the above vacuum energy is noted to be the same as in Ref.~\cite{Bellucci:2009hh}\footnote{Note that Ref.~\cite{Bellucci:2009hh} investigated the general case for the general number of extra dimensions. To compare with ours, we can take one extra dimension in their result.}, which discussed the system in the framework of standard field theory preserving the Lorentz symmetry. Note that the Casimir energy under the effects of the extra dimensions in time-like vector case can be obtained when we investigate the space-like vector case with preserve Lorentz symmetry. This will be explored in the next subsection.

\subsection{Space-like vector case in $x^{3}$-direction} 
\label{CasEnergySpacelike}
In this subsection, we turn to consider the space-like vector case $u^{(3)}=(0,0,0,1,0)$ in $x^{3}$-direction, where we have the positive- and negative-frequency solutions as follows:
\begin{eqnarray}
\psi^{(3,+)}_\beta&=&{\cal N}_\beta e^{-i\omega t} 
\begin{pmatrix}
\chi^{(3,+)}_1\\
{(-i\sigma^j\partial_j+\partial_5+i\lambda \sigma^3\partial_3)\chi^{(3,+)}_1 / (\omega+m)}
\end{pmatrix},\\
\psi^{(3,-)}_\beta&=&{\cal N}_\beta e^{i\omega t}
\begin{pmatrix}
{(i\sigma^j\partial_j-\partial_5-i\lambda \sigma^3\partial_3) \chi^{(3,-)}_2/ (\omega+m)}\\
\chi^{(3,-)}_2
\end{pmatrix}, 
\end{eqnarray}
where the two component spinors are given by the following:
\begin{eqnarray}
\chi^{(3,+)}_1&=&e^{i{\bf k}_\parallel \cdot {\bf x}_\parallel} (\phi^{(3)}_{+}e^{ik_{3}x^{3}}+\phi^{(3)}_{-}e^{-ik_{3}x^{3}}),\\
\chi^{(3,-)}_2&=&e^{-i{\bf k}_\parallel \cdot {\bf x}_\parallel} (\varphi^{(3)}_{+}e^{ik_{3}x^{3}}+\varphi^{(3)}_{-}e^{-ik_{3}x^{3}}).
\end{eqnarray}
In this space-like vector case, the eigenfrequencies read
\begin{eqnarray}
\omega=\sqrt{m^2+k^2_1+k^2_2+(1-\lambda)^{2}k^2_{3}+k^2_5}.
\end{eqnarray}

Next, we consider the behavior of the momenta in the presence of boundary conditions. At the first boundary surface $x^{3}=0$, the normal surface is given by $n_a=(0,0,0,1,0)$. At this surface, by imposing the MIT bag boundary conditions \eqref{MITBagBC}, the following equations can be obtained:
\begin{eqnarray}
\phi^{(3)}_+&=& -{m(\omega+m)+(1-\lambda)^2k^2_{3}-\sigma^3(1-\lambda)k_3(ik_5+\sigma^1k_1+\sigma^2k_2)\over (m-i(1-\lambda)k_{3})(\omega+m)}\phi^{(3)}_-,
\label{phiz1}\\
\varphi^{(3)}_-&=& -{m(\omega+m)+(1-\lambda)^2k^2_{3}-\sigma^3(1-\lambda)k_3(-ik_5+\sigma^1k_1+\sigma^2k_2)\over (m+i(1-\lambda)k_{3})(\omega+m)}\varphi^{(3)}_+,
\label{varphiz1}
\end{eqnarray}
whereas at the surface of the second plate $x^{3}=a$ with the normal surface given by $n_a=(0,0,0,-1,0)$, we have
\begin{eqnarray}
\phi^{(3)}_+&=& -{m(\omega+m)+(1-\lambda)^2k^2_{3}-\sigma^3(1-\lambda)k_3(ik_5+\sigma^1k_1+\sigma^2k_2)\over (m+i(1-\lambda)k_{3})(\omega+m)}e^{-ik_3a}\phi^{(3)}_-,
\label{phiz2}\\
\varphi^{(3)}_-&=& -{m(\omega+m)+(1-\lambda)^2k^2_{3}-\sigma^3(1-\lambda)k_3(-ik_5+\sigma^1k_1+\sigma^2k_2)\over (m-i(1-\lambda)k_{3})(\omega+m)}e^{-ik_3a}\varphi^{(3)}_+.
\label{varphiz2}
\end{eqnarray}
Comparing $\phi^{(3)}_\pm$ in Eqs.~\eqref{phiz1} and \eqref{phiz2}, we arrive at the condition where the momentum $k_{3}$ must satisfy the constraint in the form of transcendental equation as follows
\begin{eqnarray}
ma \sin(k_{3} a)+ (1-\lambda) k_{3} a \cos(k_{3}a)=0,
\label{ConstraintMomentumlam}
\end{eqnarray}
so that the momentum $k_{3}$ must be descretized. In this case, we will use $k_{n}(\equiv k_{3} a)$ as the solution of the above constraint. In the massless case, we have the same constraint that of the time-like vector case, in which we have $\cos(k_3 a)=0$, so that the solution is also given by $k_n=(n-1/2)\pi$. As noted in the time-like vector case, the same constraint will be obtained when one equals two-component spinors of negative frequency solution $\varphi^{(t)}_\pm$ in Eqs.~\eqref{varphiz1} and \eqref{varphiz2}. The above constraint shows that the solution $k_{n}$ depends on the Lorentz violation. Such a feature has also been noticed in Refs.~\cite{Cruz:2018thz, Rohim:2023tmy}. Under the quasiperiodic boundary conditions, we find that in this space-like vector case, the momentum $k_5$ is also discretized with a similar form  to Eq.~\eqref{k5discrete}. Thus, the eigenfrequency is given by
\begin{eqnarray}
\omega_{n,\ell}=\sqrt{m^2+k^2_1+k^2_2+\left({(1-\lambda)k_{n}\over a}\right)^2+k^2_{\ell}}.
\end{eqnarray}

The field expansion for the case of Lorentz violation in $x^3$-direction is given by
\begin{eqnarray}
    \Psi^{(3)}=\int dk_1\int dk_2\sum_{n=1}^\infty \sum_{\ell=-\infty}^\infty \sum_{s=1}^2\bigg[ b_{k_1,k_2,n,\ell, s} \psi^{(3,+)}_{k_1,k_2, n,\ell, s} +d^{\dagger}_{k_1,k_2,n,\ell, s}  \psi^{(3,-)}_{k_1,k_2, n,\ell, s} \bigg], 
\label{fieldexpx3}
\end{eqnarray}
where the positive- and negative-frequency solutions satisfy the orthonormality conditions as given in Eq.~\eqref{othonormalitytime}. The Hamiltonian for the case of Lorentz violation in $x^3$-direction is given by
\begin{eqnarray}
    \hat H=\int d{\bf x}_\parallel \int_0^a dx^3 \bar\Psi^{(3)}(-i\gamma^j\partial_j+m+i\lambda \gamma^3\partial_3)\Psi^{(3)}=i\int d{\bf x}_\parallel \int_0^a dx^3 \Psi^{(3)\dagger}\partial_t\Psi^{(3)}.   \end{eqnarray}
Straightforwardly, performing a similar way as in the time-like vector case, we can obtain the vacuum energy as follows,
\begin{eqnarray}
E_{\rm Vac.}=-{L^2\over 2\pi^2}\int {d k_1}\int {d k_2}\sum_{\ell=-\infty}^\infty\sum^\infty_{n=1}\sqrt{m^2+k^2_1+k^2_2+\left({(1-\lambda)k_{n}\over a}\right)^2+k^2_{\ell}}.
\end{eqnarray}
From the above expression, it can be observed that both Lorentz violation in $x^{3}$-direction and the presence of the compact dimensions have an impact on the vacuum energy of the Dirac field.  It is also important to note that the calculation for the Casimir energy in the case of $x^{1}$- and $x^{2}$- directions is rather trivial, since they lead to the vacuum energy of the time-like vector case \eqref{time-like-vac} multiplied by a factor $1/(1-\lambda)$ (see also Ref.~\cite{Cruz:2018thz}). In Appendix~\ref{casex12}, we have derived these calculations and shown that they are not relevant to the purpose of this study.

Below, we will derive the Casimir energy from the above vacuum energy. In fact,
 one can see that the above vacuum state is divergent. To tackle this issue, we will use the Abel-Plana-like summation given as follows \cite{Romeo:2000wt}
\begin{eqnarray}
\sum_{n=1}^{\infty} {\pi f(k_{n})\over \left(1-{\sin(2 k_{n})\over 2 k_{n}}\right)}=-{\pi mb f(0)\over 2(m b+1)}+\int_0^\infty dz f(z)-i\int_0^\infty dt {f(it)-f(-it)\over {t+mb\over t-mb}e^{2t}+1}, 
\label{AbelPlanaLike}
\end{eqnarray}
where
\begin{eqnarray}
b={a\over (1-\lambda)}.
\label{defb}
\end{eqnarray}
From the momentum constraint \eqref{ConstraintMomentumlam}, we have the relation for the denominator of the left-hand side of Eq.~\eqref{AbelPlanaLike} as follows
\begin{eqnarray}
1-{\sin(2 k_n)\over 2 k_n}=1+{mb\over (mb)^2+k^2_n}.
\end{eqnarray}
Then, the vacuum energy reads
\begin{eqnarray}
E_{\rm Vac.}=-{L^2\over 2\pi^3 b} \int dk_1\int dk_2\sum_{\ell=-\infty}^\infty\left(-{\pi mb f(0)\over 2(mb+1)}+\int_0^\infty dz f(z)-i\int_0^\infty dt {f(it)-f(-it)\over {t+mb\over t-mb}e^{2t}+1}\right), 
\end{eqnarray}
where 
\begin{eqnarray}
f(z)=\sqrt{k^2_1 b^2+k^2_2b^2+z^2+m^2_\ell b^2} \left(1+{mb\over (mb)^2+k^{2}_{n}}\right),
\end{eqnarray}
with $m^2_\ell=m^2+k^2_{\ell}$ explicitly given by
\begin{eqnarray}
m^2_\ell=m^2+[2\pi(\ell+\beta)/q]^2.
\end{eqnarray}

We can separate the above vacuum energy  into three parts as follows
\begin{eqnarray}
E_{\rm Vac.}=bE^{(0)}_{\rm Vac.}+2E^{(1)}_{\rm Vac}+\Delta E_{\rm Vac.}
\end{eqnarray}
The first, second, and third parts correspond to vacuum energy in the absence of two plates, in the presence of a single plate, and interaction, respectively. The vacuum energy $E^{(0)}_{\rm Vac.}$ is explicitly given by the following:
\begin{eqnarray}
E^{(0)}_{\rm Vac.}=-{L^2\over  4\pi^{3}}\int d k_1 \int dk_2\int dk_3\sum_{\ell=-\infty}^\infty\sqrt{m^2_\ell+k^2_1+k^2_2+k^2_3},
\end{eqnarray}
whereas the vacuum energy $E^{(1)}_{\rm Vac.}$ is given by the following:
\begin{eqnarray}
    E^{(1)}_{\rm Vac.}=-{L^2\over  4\pi^{3}}\int dk_1\int dk_2\sum_{\ell=-\infty}^\infty\bigg(\sqrt{m^2_\ell+k^2_1+k^2_2}+\int_0^\infty dk_{3}{m\sqrt{m^2_\ell+k^2_1+k^2_2+k^2_3}\over m^2+k_{3}^2}\bigg).
\end{eqnarray}
From the above expression, one can see that the vacuum energy $E^{(1)}_{\rm Vac.}$ does not depend on the plates' distance. Furthermore, the Casimir energy can be understood as the difference between the vacuum energy in the presence of plates and that in the absence of one. Therefore, we can use the interaction part of the above vacuum energy as the Casimir energy of the Dirac field in the presence of two parallel plates ($E_{\rm Cas.}\equiv \Delta E_{\rm Vac.} $). 

The Casimir energy  is given by the last term of Eq.~\eqref{AbelPlanaLike} as follows
\begin{eqnarray}
E_{\rm Cas.}&=&{iL^2b \over 2\pi^3}\int dk_1\int dk_2\sum_{\ell=-\infty}^\infty \int_0^\infty du {u-m\over (u+m)e^{2bu}+u-m} \left(1+{mb\over (mb)^2-(ub)^2}\right) \nonumber\\
&&~~\times \bigg[\sqrt{k^2_1+k^2_2+(iu)^2+m^2_\ell} - \sqrt{k^2_1+k^2_2+(-iu)^2+m^2_\ell} \bigg], 
\end{eqnarray}
where we have introduced a new variable $t=bu$. In the above expression, we can separate the integration over $u$ into two parts or intervals, i.e., $[0, \sqrt{k_1^2+k_2^2+m^2_\ell]}$ and  $[\sqrt{k_1^2+k_2^2+m^2_\ell}, \infty]$. The first part of the $u$ integral vanishes whereas the second one does not. Thus, the Casimir energy reads 
\begin{eqnarray}
E_{\rm Cas.}= -{L^2 \over \pi^3}\int dk_1\int dk_2\sum_{\ell=-\infty}^\infty \int_{\sqrt{k_1^2+k_2^2+m_\ell}}^\infty du {(u-m) b-m/(u+m)\over (u+m)e^{2bu}+u-m} \sqrt{u^2-{\bf k}^2_p-m^2_\ell}. 
\end{eqnarray}
To further proceed, we next use the following formula \cite{Bellucci:2009hh}
\begin{eqnarray}
\int d {\bf k}_p \int^\infty_{\sqrt{{\bf k}^2_p
+c^2}} du (u^2-{\bf k}^2_p-c^2)^{(s+1)/2}f(u)={\pi^{p/2} \Gamma[{s+3\over 2}]\over \Gamma[{p+s+3\over 2}]}\int_c^\infty dz (z^2-c^2)^{(p+s+1)/2} f(z),
\label{FormulaInt}
\end{eqnarray}
where in our case, $p=2$ and $s=0$, so that the Casimir energy now reads as follows:
\begin{eqnarray}
E_{\text{Cas.}}=-{(4\pi)^{-3/2}4 L^2\over \Gamma(5/2)}\sum_{\ell=-\infty}^\infty \int_{m_\ell}^\infty dz {(z^2-m^2_\ell)^{3/2}\over (z+m)e^{2 b z}+z-m}\bigg(b (z-m)-{m\over z+m}\bigg) . 
\label{eq:casme}
\end{eqnarray}
Rewriting the last factor of Eq.~\eqref{eq:casme} using the following relations;  
\begin{eqnarray}
{b (z-m)-m/ (z+m)\over  (z+m)e^{2 b z}+z-m}=-{1\over 2}{d\over dz}\ln\bigg(1+{z-m\over z+m}e^{-2 b z}\bigg),
\end{eqnarray}
we obtain that
\begin{eqnarray}
E_{\rm Cas.}=-{L^2\over \pi^2}\sum_{\ell=-\infty}^\infty \int_{m_\ell}^\infty dz z (z^2-m^2_\ell)^{1/2} \ln\bigg(1+{z-m\over z+m}e^{-2 b z}\bigg),
\label{ECas}
\end{eqnarray}
where we have performed integration by part. 
By introducing a new variable $bz = x +bm_\ell$, the Casimir energy can be rewritten as follows
\begin{eqnarray}
   E_{\rm Cas.}=-{L^2\over b^3\pi^2}\sum_{\ell=-\infty}^{+\infty}\int_0^\infty dx (x+bm_\ell) (x^2+2xbm_\ell)^{1/2} \ln\bigg(1+{x+bm_\ell-bm\over x+bm_\ell+bm}e^{-2 (x+bm_\ell)}\bigg).
\end{eqnarray}
Without an extra dimension, this Casimir energy reduces to that in Ref.~\cite{Cruz:2018thz}.

\begin{figure}[tbp]
\centering 
\includegraphics[width=.49\textwidth]{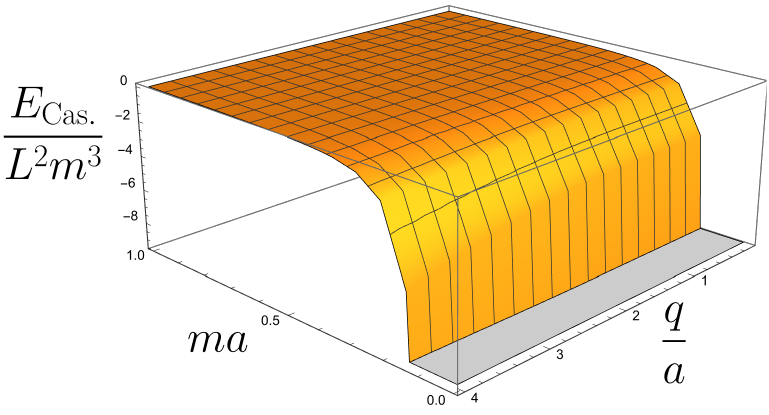}
\hfill
\includegraphics[width=.49\textwidth]{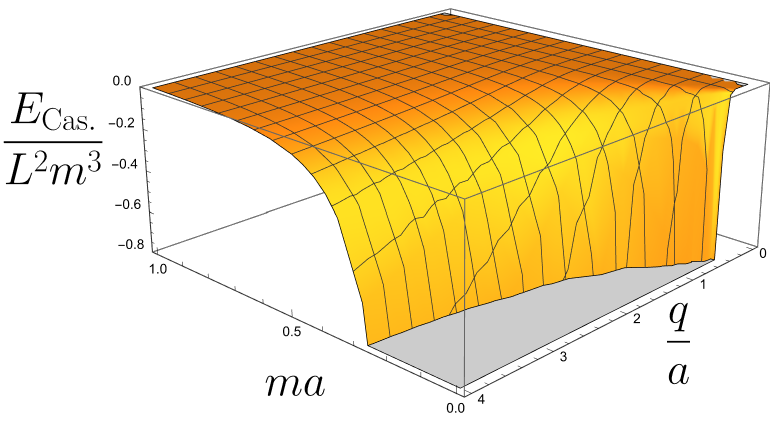}
\caption{\label{ECas3Dalpha} Plot of the scaled Casimir energy ${E}_{\rm Cas.}/(L^2m^3)$ as a function of $ma$ and $q/a$  for two values of parameter $\beta$ with a fixed value of Lorentz violation intensity $\lambda=0.1$. In the left panel, we use $\beta=0$ while in the right panel, we use $\beta=0.5$. This figure shows the Lorentz violation in the $x^3$-direction.}
\end{figure}

\begin{figure}[h!]
\centering 
\includegraphics[height=.3\textwidth]{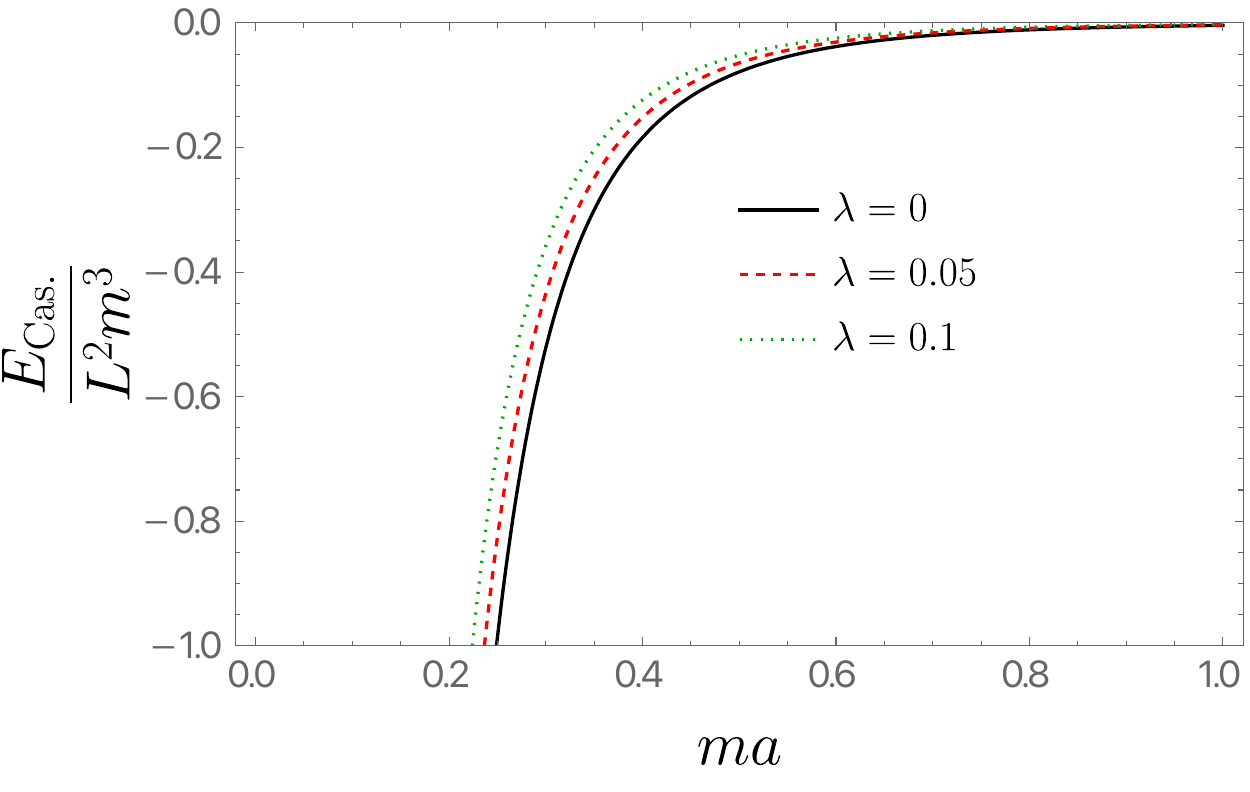}
\hfill
\includegraphics[height=.3\textwidth]{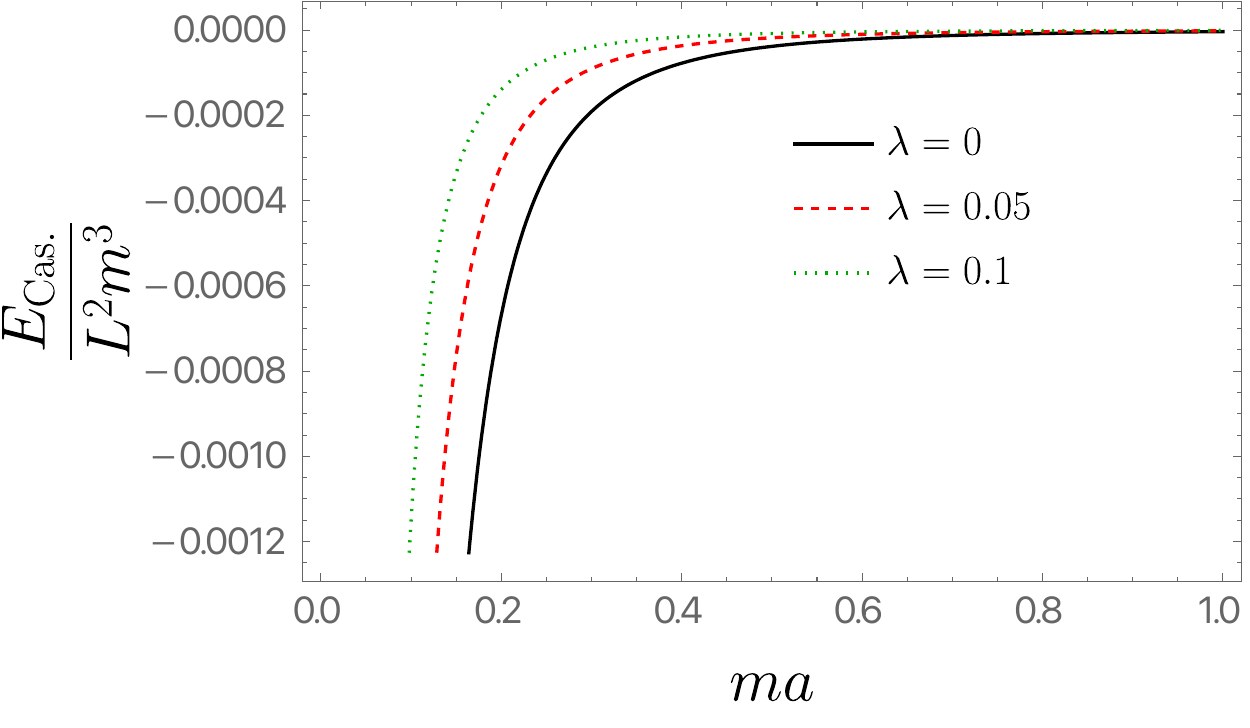}
\caption{\label{SCEma} Plot of the scaled Casimir energy ${E}_{\rm Cas.}/(L^2m^3)$ as a function of $ma$ for various values of the Lorentz violation's intensity $\lambda=0,0.05,0.1$ with fixed $q/a=0.5$ and two values of $\beta$. In the left panel, we use $\beta=0$ while in the right panel, we use $\beta=0.5$. This figure shows the Lorentz violation in the $x^3$-direction.}  
\end{figure}

\begin{figure}[tbp]
\centering 
\includegraphics[width=.49\textwidth]{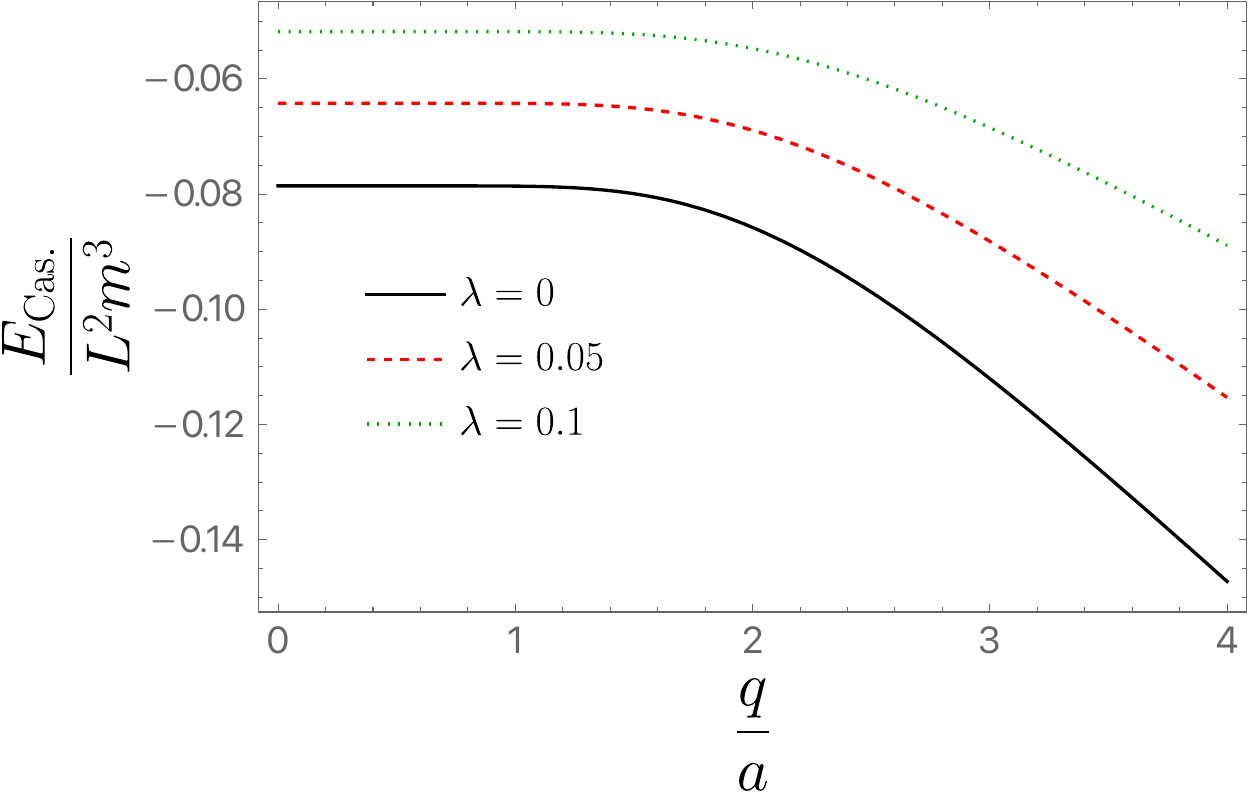}
\hfill
\includegraphics[width=.49\textwidth]{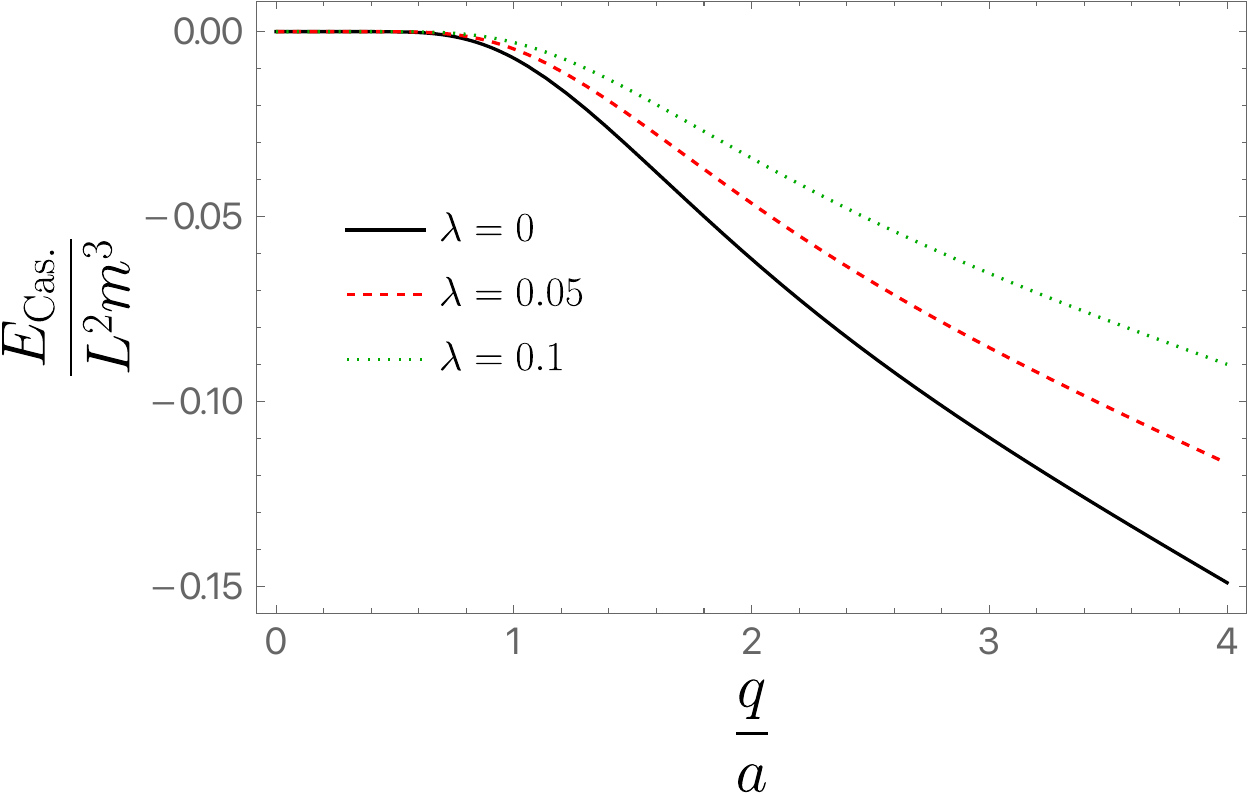}
\caption{\label{SCEqa} Plot of the scaled Casimir energy ${E}_{\rm Cas.}/(L^2m^3)$ as a function of $q/a$ for various values of the Lorentz violation's intensity $\lambda=0,0.05,0.1$ with fixed $m/a=0.5$ and two values of $\beta$. In the left panel, we use $\beta=0$ while in the right panel, we use $\beta=0.5$. This figure shows the Lorentz violation in the $x^3$-direction. 
}
\end{figure} 

\begin{figure}[tbp]
\centering 
\includegraphics[width=.49\textwidth]{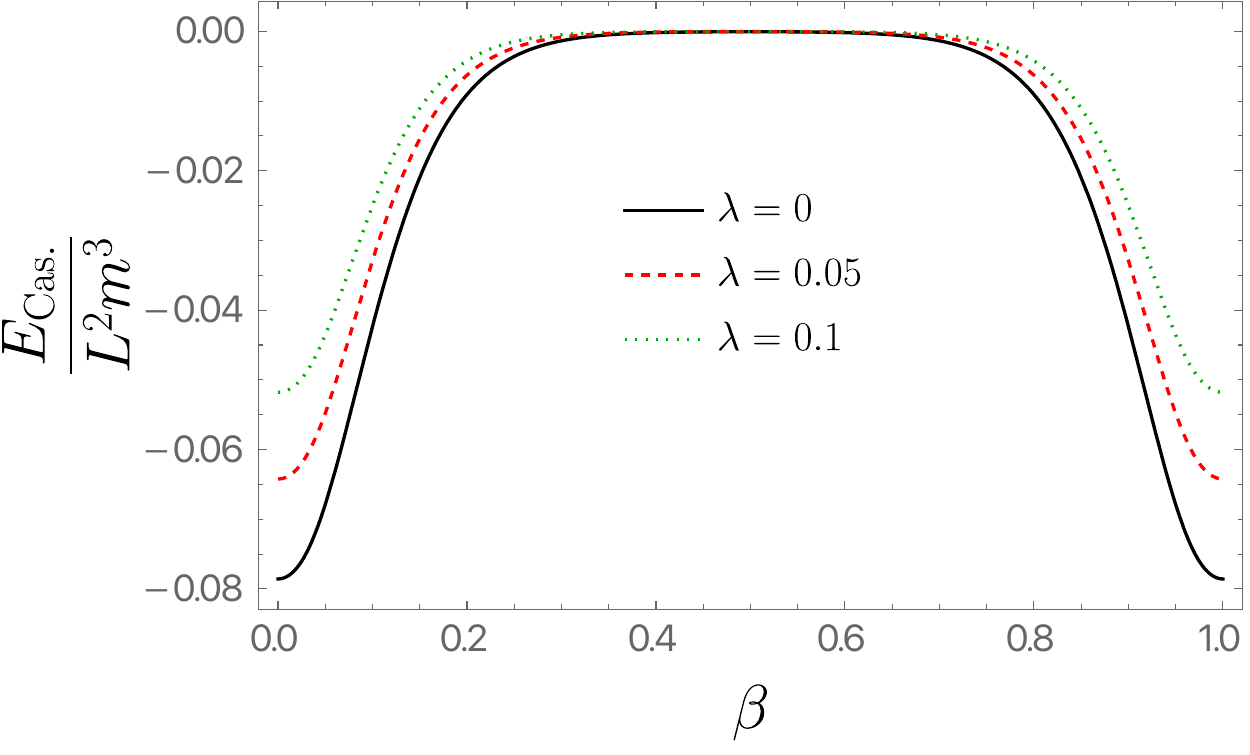}
\hfill
\includegraphics[width=.49\textwidth]{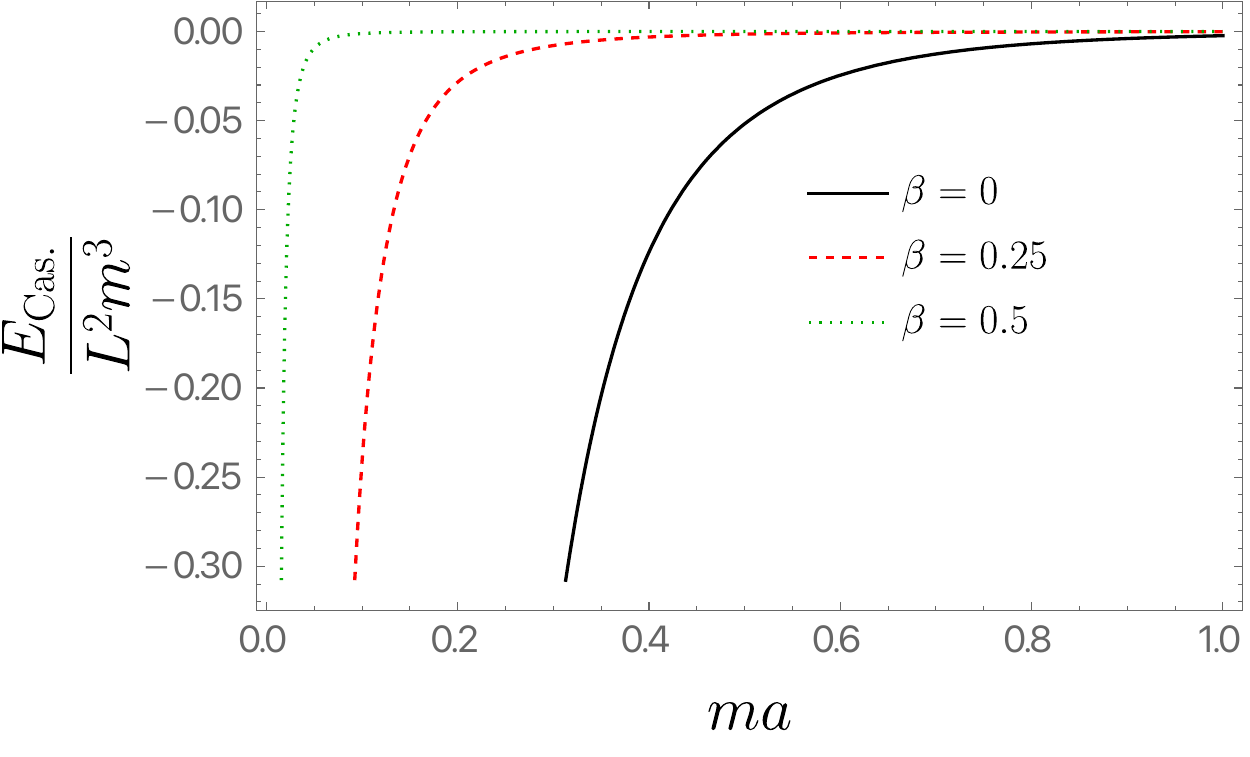}
\caption{ \label{ECasalpha}
The left panel shows the scaled Casimir energy  ${E}_{\rm Cas.}/(L^2m^3)$ as a function of parameter $\beta$ for three values of Lorentz violation's intensity $\lambda=0, 0.05,0.1$ with $\beta=0$ whereas the right panel shows the scaled Casimir energy as a function of $ma$ for various values of  $\beta=0,0.25, 0.5$ with fixed $q/a=0.5$ and $\lambda=0.1$. This figure shows the Lorentz violation in $x^3$-direction.}
\end{figure} 
 
Figure \ref{ECas3Dalpha} shows the behavior of the scaled Casimir energy  ${E}_{\rm Cas.}/(L^2m^3)$ as a function of $ma$ and $q/a$ for two various values of $\beta$ with a fixed value of the Lorentz violation parameter $\lambda=0.1$. The left panel is for $\beta=0$ while the right one is for $\beta=0.5$. From this figure, it can be seen that the Casimir energy goes to zero as the parameter $ma$ increases. In particular, the Casimir energy in the case of $\beta=0.5$ goes to zero faster than in the case of $\beta=0$. The behavior of this figure is favored by Figs.~\ref{SCEma} and \ref{SCEqa}. Moreover, one can see that as the parameter $\lambda$ increases,  the magnitude of the Casimir energy decreases. For the case of $\beta=0$, the plot of the Casimir energy as a function of $q/a$ tends to have constant value in the region of $q/a <1$ because the zero-mode of extra dimension dominates (see the left panel of Fig.~\ref{SCEqa}). If the value of $q/a$ is sufficiently large, the other modes contribute so that the magnitude Casimir energy increases. As for $\beta=0.5$ and within a small value of $ma$, the magnitude of the scaled Casimir energy increases as the value $q/a$ increases (see right panel of Fig.~\ref{SCEqa}). The magnitude of the Casimir energy is symmetric under the changing of parameter $\beta$, in which the maximum amplitude of the Casimir energy is achieved when $\beta=0$, whereas the minimum one is achieved when $\beta=0.5$ (see Fig.~\ref{ECasalpha}). We note that the latter feature is consistent with that of Ref.~\cite{deFarias:2023xjf} in the case of the EM field.

Below we discuss the Casimir energy for certain limits. Taking massless limit from Eq.~\eqref{ECas}, the Casimir energy reads
\begin{eqnarray}
E_{\rm Cas.}=-{L^2\over \pi^2}\sum_{\ell=-\infty}^\infty \int_{k_\ell}^\infty dz z (z^2-k^2_\ell)^{1/2} \ln\bigg(1+e^{-2 b z}\bigg).
\end{eqnarray}
In the case of $q\gg a$ for the Casimir energy \eqref{ECas}, the dominant contributions come from large $\ell$. Then we can replace the summation $\sum_{\ell=-\infty}^{+\infty}$ as integral $\int d \ell$ (see Ref.~\cite{Bellucci:2009hh}). This argument can be easily shown as follows. We consider the summation of a function $f (\ell)$ that is given by,
\begin{eqnarray}
  \sum_{\ell = - \infty}^{+ \infty} f (\ell) & = & \lim_{N \rightarrow \infty}
  \sum_{\ell = - N}^{+ N} f (\ell) N \frac{\Delta \ell}{N},
\end{eqnarray}
where $\Delta \ell(\equiv \ell_{j+1}-\ell_j)$ represents the discrete interval between consecutive $\ell$-values.  We can define $\ell = N r$. Since $N$ goes to infinity, then $(\Delta \ell /
N) \equiv d r$ becomes infinitesimal and we may replace the summation as an
integral. Therefore, the above expression changes into
\begin{eqnarray}
  \sum_{\ell = - \infty}^{+ \infty} f (\ell) & = & \int_{r = -
  \infty}^{\infty} f (N r) d (N r) = \int_{\ell = - \infty}^{\infty} f (\ell) d \ell .
\end{eqnarray}
Now the Casimir energy reads as follows:
\begin{eqnarray}
E_{\rm Cas.}\simeq-{L^2\over \pi^2}\int d \ell\int_{m_\ell}^\infty dz z (z^2-m^2_\ell)^{1/2} \ln\bigg(1+{z-m\over z+m}e^{-2 b z}\bigg). \label{casen}
\end{eqnarray}
By using the similar formula given in Eq.~\eqref{FormulaInt}, the above integral expression can be more explicitly written as follows 
\begin{eqnarray}
\label{FormulaIntx3}
\int d\ell\int_{\sqrt{k^2_{\ell}+m^2}} dz (z^2-k^2_{\ell}-m^2)^{1/2}f(z)= 
{q\over 4}
\int^\infty_m dz (z^2-m^2)f(z),
\end{eqnarray}
where we have used $d \ell=(q/2\pi)dk_\ell$, so that the Casimir energy in the case of $q \gg a$ reads 
\begin{eqnarray}
E_{\rm Cas.}=-{q L^2\over 4\pi^2}\int^\infty_m dzz(z^2-m^2)  \ln\bigg(1+{z-m\over z+m}e^{-2 b z}\bigg). 
\end{eqnarray}
Taking the massless limit from the above equation, we obtain
\begin{eqnarray}
E_{\rm Cas.}=-{45 qL^2 \zeta(5)\over 512\pi^2 b^4} = -{45 qL^2 \zeta(5) (1-\lambda)^{4} \over 512\pi^2 a^4}, \label{Ecasm0x3}
\end{eqnarray}
where $\zeta (x)$ is the Riemann zeta function. We next consider the Casimir energy in the case of $q \ll a$ and $\beta=0$. In this case, we have
\begin{eqnarray}
E_{\rm Cas.}=-{L^2\over \pi^2}\int_m^\infty dz z (z^2-m^2)^{1/2} \ln\bigg(1+{z-m\over z+m}e^{-2 b z}\bigg).
\label{EqECasx31}
\end{eqnarray}
Introducing new variable $bz=y+bm$, we have
\begin{eqnarray}
E_{\rm Cas.}=-{L^2\over \pi^2 b^3}\int_0^\infty dy (y+bm) (y^2+2ybm)^{1/2} \ln\bigg(1+{y\over y+2bm}e^{-2 (y+bm)}\bigg), 
\label{EqECasx32}
\end{eqnarray}
where we recover the result by Ref.~\cite{Cruz:2018thz} which is discussed in the usual 1+3 Minkowski spacetime. In the massless case together with $q \ll a$ with $\beta=0$, the above Casimir energy reduces to $E_{\rm Cas.}=7 \pi^2 L^2/2880 b^3$.

\subsection{Space-like vector case in $x^5$-direction }

Let us consider the Casimir effect for the space-like vector case $u^{(5)}=(0,0,0,0,1)$ in $x^5$-direction. For this purpose, the field representations of both positive and negative frequencies are written as follows,
\begin{eqnarray}
\psi^{(5,+)}_\beta&=&{\cal N}_\beta e^{-i\omega t} 
\begin{pmatrix}
\chi^{(5,+)}_1\\
{(-i\sigma^j\partial_j+(1-\lambda)\partial_5)} \chi^{(5,+)}_1/(\omega+m)
\end{pmatrix},\\
\psi^{(5,-)}_\beta&=&{\cal N}_\beta e^{i\omega t}
\begin{pmatrix}
{(i\sigma^j\partial_j-(1-\lambda)\partial_5) \chi^{(5,-)}_2 / (\omega+m)}\\
\chi^{(5,-)}_2
\end{pmatrix}, 
\end{eqnarray}
respectively, where the eigenfrequency is given by
\begin{eqnarray}
\omega=\sqrt{m^2+k_1^2+k_2^2+k^2_3+(1-\lambda)^2k^2_5}
\end{eqnarray}
and the two-component spinors are given by
\begin{eqnarray}
\chi^{(5,+)}_1&=&e^{i{\bf k}_{\parallel} \cdot {\bf z}_{\parallel}} (\phi^{(5)}_{+}e^{ik_{3}z^{3}}+\phi^{(5)}_{-}e^{-ik_{3}z^{3}}),\\
\chi^{(5,-)}_2&=&e^{-i{\bf k}_\parallel \cdot {\bf z}_\parallel}  (\varphi^{(5)}_{+}e^{ik_{3}z^{3}}+\varphi^{(5)}_{-}e^{-ik_{3}z^{3}}).
\end{eqnarray}

At the first boundary surface $x^{3}=0$, the normal surface is given by $n_a=(0,0,0,+1,0)$. At this surface, by imposing the MIT bag boundary conditions, the following equations are obtained:
\begin{eqnarray}
\phi^{(5)}_+&=& -{m(\omega+m)+k^2_{3}-\sigma^3 k_3(i(1-\lambda)k_5+\sigma^1k_1+\sigma^2k_2)\over (m-ik_{3})(\omega+m)}\phi^{(5)}_-,
\label{phi51}\\
\varphi^{(5)}_-&=& -{m(\omega+m)+k^2_{3}-\sigma^3 k_3(-i(1-\lambda)k_5+\sigma^1k_1+\sigma^2k_2)\over (m+ik_{3})(\omega+m)}\varphi^{(5)}_+,
\label{varphi51}
\end{eqnarray}
whereas at the surface of the second plate $x^{3}=a$ with the normal surface given by $n_a=(0,0,0,-1,0)$, we have
\begin{eqnarray}
\phi^{(5)}_+&=& -{m(\omega+m)+k^2_{3}-\sigma^3 k_3(i(1-\lambda)k_5+\sigma^1k_1+\sigma^2k_2)\over (m+ik_{3})(\omega+m)}e^{-ik_3a}\phi^{(5)}_-,
\label{phi52}\\
\varphi^{(5)}_-&=& -{m(\omega+m)+k^2_{3}-\sigma^3 k_3(-i(1-\lambda)k_5+\sigma^1k_1+\sigma^2k_2)\over (m-i k_{3})(\omega+m)}e^{-ik_3a}\varphi^{(5)}_+.
\label{varphi52}
\end{eqnarray}
We note that the momentum $k_3$ satisfies the same constraint as in Eq.~\eqref{ConstraintMomentum},
so that the momentum $k_3$ here must also be discretized. We will use $k_{n}(\equiv k_{3} a)$ as the solution to the above constraint. In this space-like case, we also note that the momentum $k_5$ is also descretized as given in Eq.~\eqref{k5discrete} so that under the boundary conditions the eigenfrequencies are then read as follows:
\begin{eqnarray}
\omega_{n,\ell}=\sqrt{m^2+k^2_1+k^2_2+\left({k_{n}\over a}\right)^2+(1-\lambda)^2k^2_{\ell}}.
\end{eqnarray}

The next task is to derive the vacuum energy with the Lorentz violation in $x^5$-direction. 
For this purpose, the field expansion can be written as follows,
\begin{eqnarray}
    \Psi^{(5)}=\int dk_1\int dk_2\sum_{n=1}^\infty \sum_{\ell=-\infty}^\infty \sum_{s=1}^2\bigg[ b_{k_1,k_2,n,\ell, s} \psi^{(5,+)}_{k_1,k_2, n,\ell, s} +d^{\dagger}_{k_1,k_2,n,\ell, s}  \psi^{(5,-)}_{k_1,k_2, n,\ell, s} \bigg], 
\label{fieldexpx5}
\end{eqnarray}
where the positive- and negative-frequency solutions in the $x^5$-direction satisfy the orthonormality conditions in Eq.~\eqref{othonormalitytime}. 
One can easily show that the Hamiltonian in the $x^5$-direction case has the following form,
\begin{eqnarray}
    \hat H=\int d{\bf x}_\parallel \int_0^a dx^3 \bar\Psi^{(5)}(-i\gamma^j\partial_j+m+i\lambda \gamma^5\partial_5)\Psi^{(5)}=i\int d{\bf x}_\parallel \int_0^a dx^3 \Psi^{(5)\dagger}\partial_t\Psi^{(5)},  
     \end{eqnarray}
and the vacuum energy is then given as follows
\begin{eqnarray}
E_{\rm Vac.}=-{L^2\over 2\pi^2}\int {d k_1}\int {d k_2}\sum_{\ell=-\infty}^\infty\sum^\infty_{n=1}\sqrt{m^2+k^2_1+k^2_2+\left({k_{n}\over a}\right)^2+(1-\lambda)^2k^2_{\ell}}.
\end{eqnarray}
Following the same procedure as we did in the space-like vector case of $x^3$ direction, straightforwardly, we can obtain the Casimir energy in the space-like vector case $x^5$-direction as follows,
\begin{eqnarray}
  E_{\rm Cas.}=-{L^2\over \pi^2}\sum_{\ell=-\infty}^{+\infty}\int_{\tilde m_\ell}^\infty dz z(z^2-\tilde m_\ell^2)^{1/2} \ln\bigg(1+{z-m\over z+m}e^{-2az}\bigg), \label{casx5}
\end{eqnarray}
where 
\begin{eqnarray}
\tilde m^2_\ell=m^2+(1-\lambda)^2k^2_\ell.
\label{defml}
\end{eqnarray}
By introducing $az=x+a\tilde m_\ell$, we have
\begin{align}
   E_{\rm Cas.}=-{L^2\over a^3\pi^2}\sum_{\ell=-\infty}^{+\infty}\int_0^\infty dx (x+a\tilde m_\ell) (x^2+2xa \tilde m_\ell)^{1/2} \ln\bigg(1+{x+a\tilde m_\ell-am\over x+a\tilde m_\ell+am}e^{-2 (x+a\tilde m_\ell)}\bigg).
   \label{EqECasx52}
\end{align}

\begin{figure}[tbp]
\centering 
\includegraphics[width=.49\textwidth]{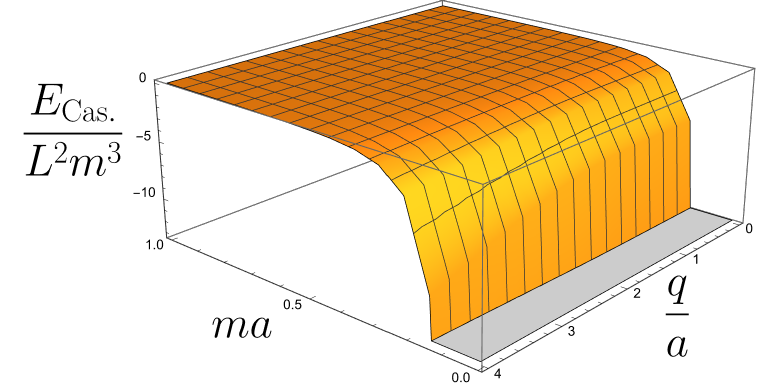}
\hfill
\includegraphics[width=.49\textwidth]{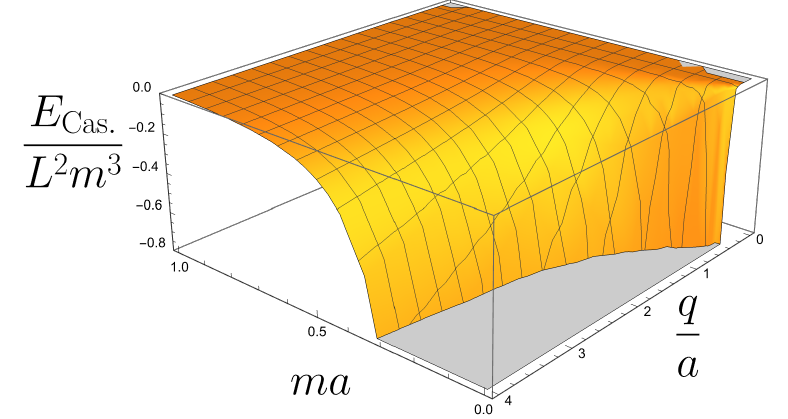}
\caption{\label{ECas3Dalphax5} Plot of the scaled Casimir energy ${E}_{\rm Cas.}/(L^2m^3)$ as a function of $ma$ and $q/a$  for two values of parameter $\beta$ with a fixed value of Lorentz violation intensity $\lambda=0.1$. In the left panel, we use $\beta=0$ whereas in the right panel, we use $\beta=0.5$. This figure shows the Lorentz violation in the $x^5$-direction.}
\end{figure}

\begin{figure}[tbp]
\centering 
\includegraphics[width=.49\textwidth]{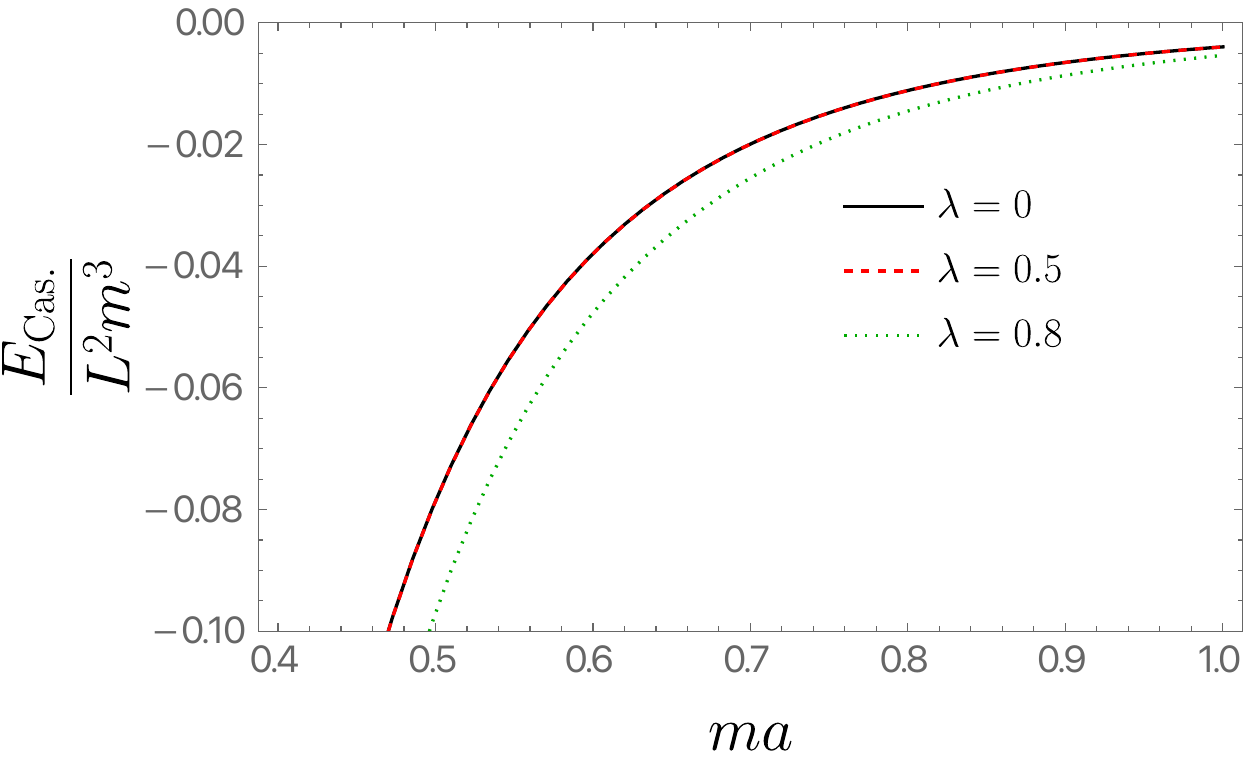}
\hfill
\includegraphics[width=.49\textwidth]{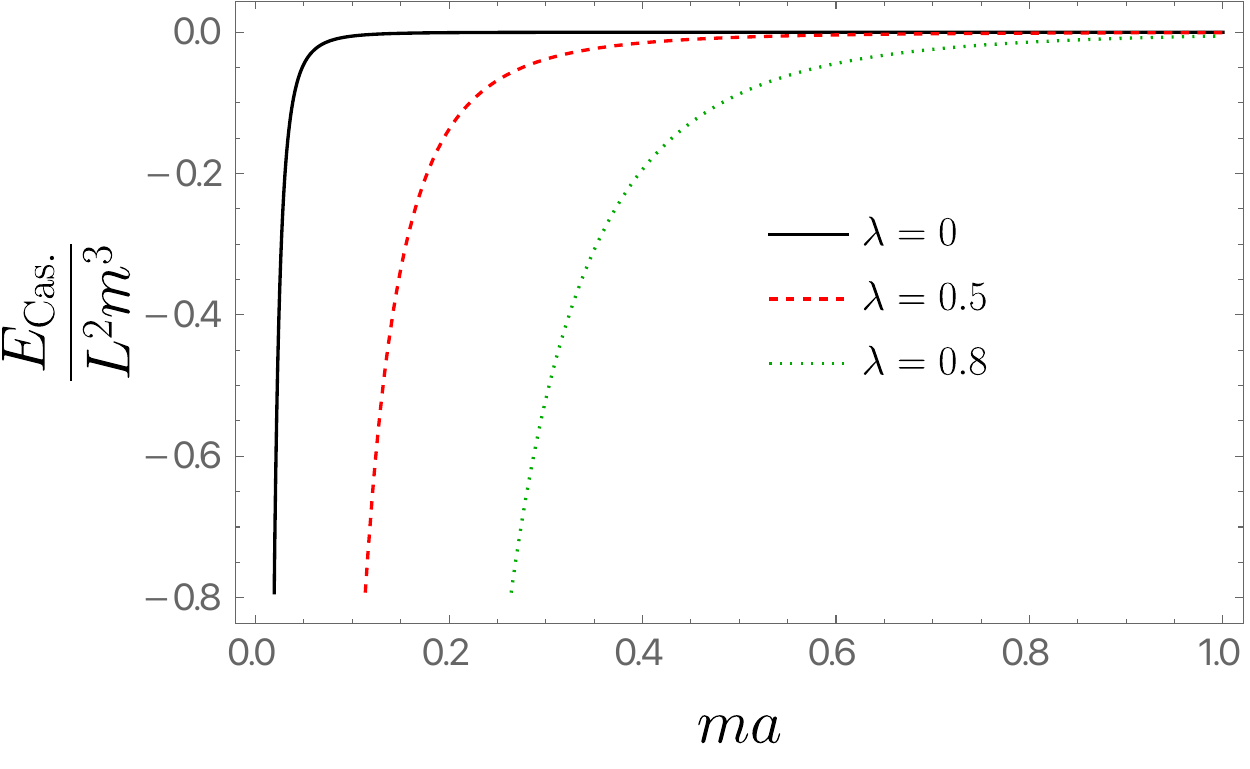}
\caption{\label{ECasmax5} Plot of the scaled Casimir energy ${E}_{\rm Cas.}/(L^2m^3)$ as a function of $ma$ for various values of the Lorentz violation's intensity $\lambda=0,0.5,0.8$ with fixed $q/a=0.5$ and two values of $\beta$. In the left panel, we use $\beta=0$ whereas in the right panel, we use $\beta=0.5$.  This figure shows the Lorentz violation in the $x^5$-direction.}
\end{figure}

\begin{figure}[tbp]
\centering 
\includegraphics[width=.49\textwidth]{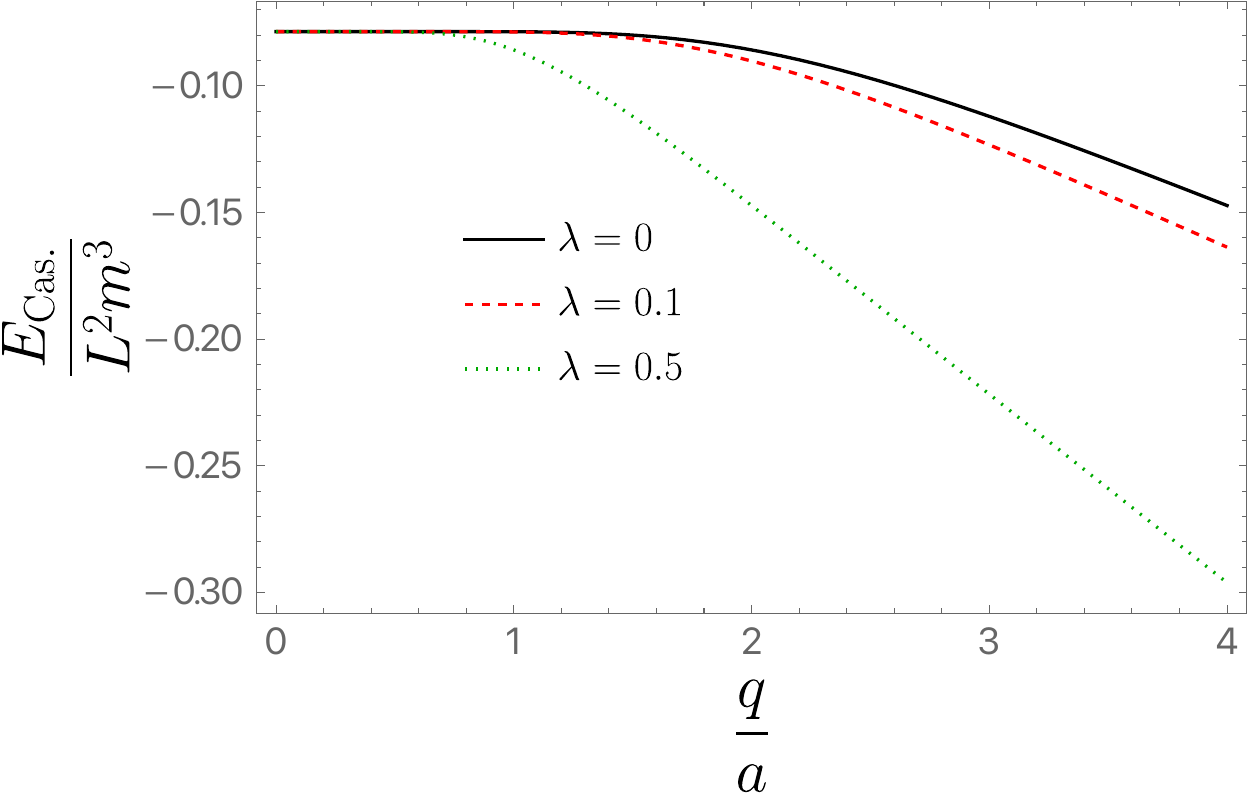}
\hfill
\includegraphics[width=.49\textwidth]{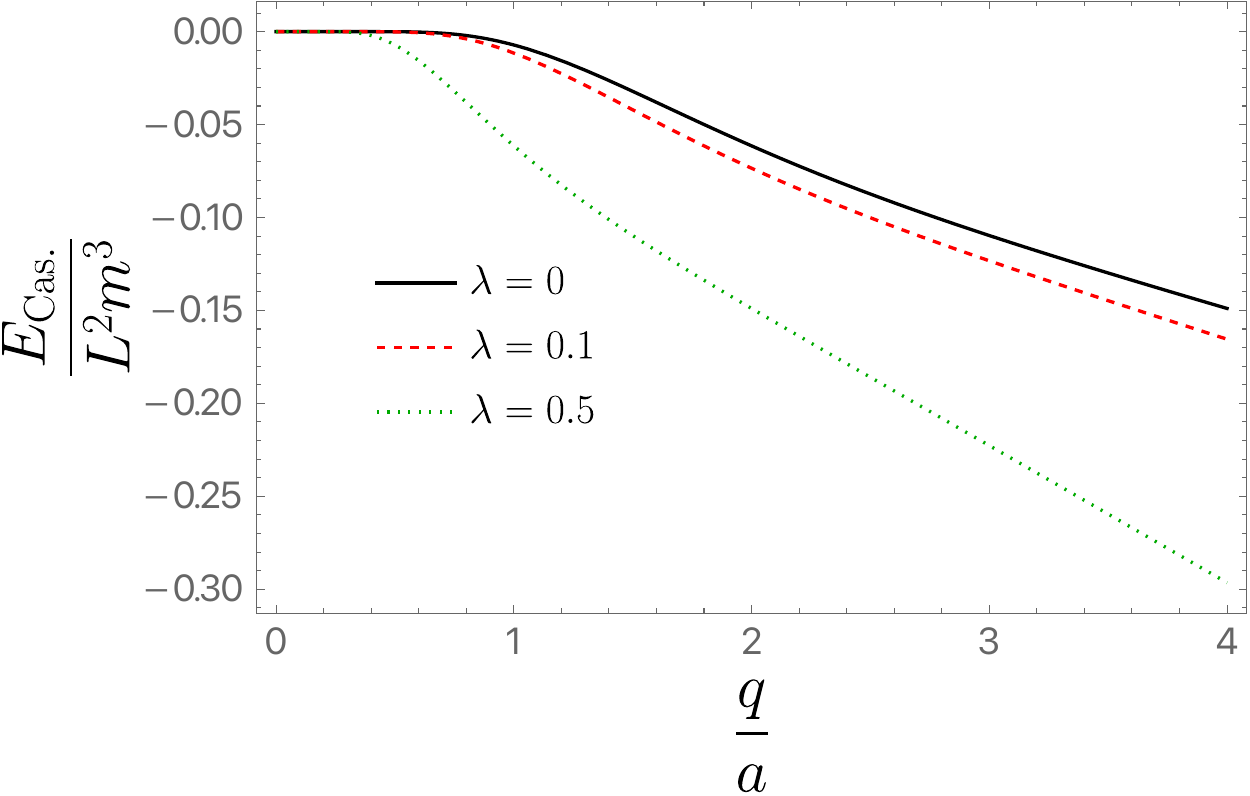}
\caption{\label{ECasqax5}Plot of the scaled Casimir energy ${E}_{\rm Cas.}/(L^2m^3)$ as a function of $q/a$ for various values of the Lorentz violation's intensity $\lambda=0,0.1,0.5$ with fixed $ma=0.5$ and two values of $\beta$. In the left panel, we use $\beta=0$ whereas in the right panel, we use $\beta=0.5$. This figure shows the Lorentz violation in the $x^5$-direction.}
\end{figure} 

\begin{figure}[tbp]
\centering 
\includegraphics[width=.49\textwidth]{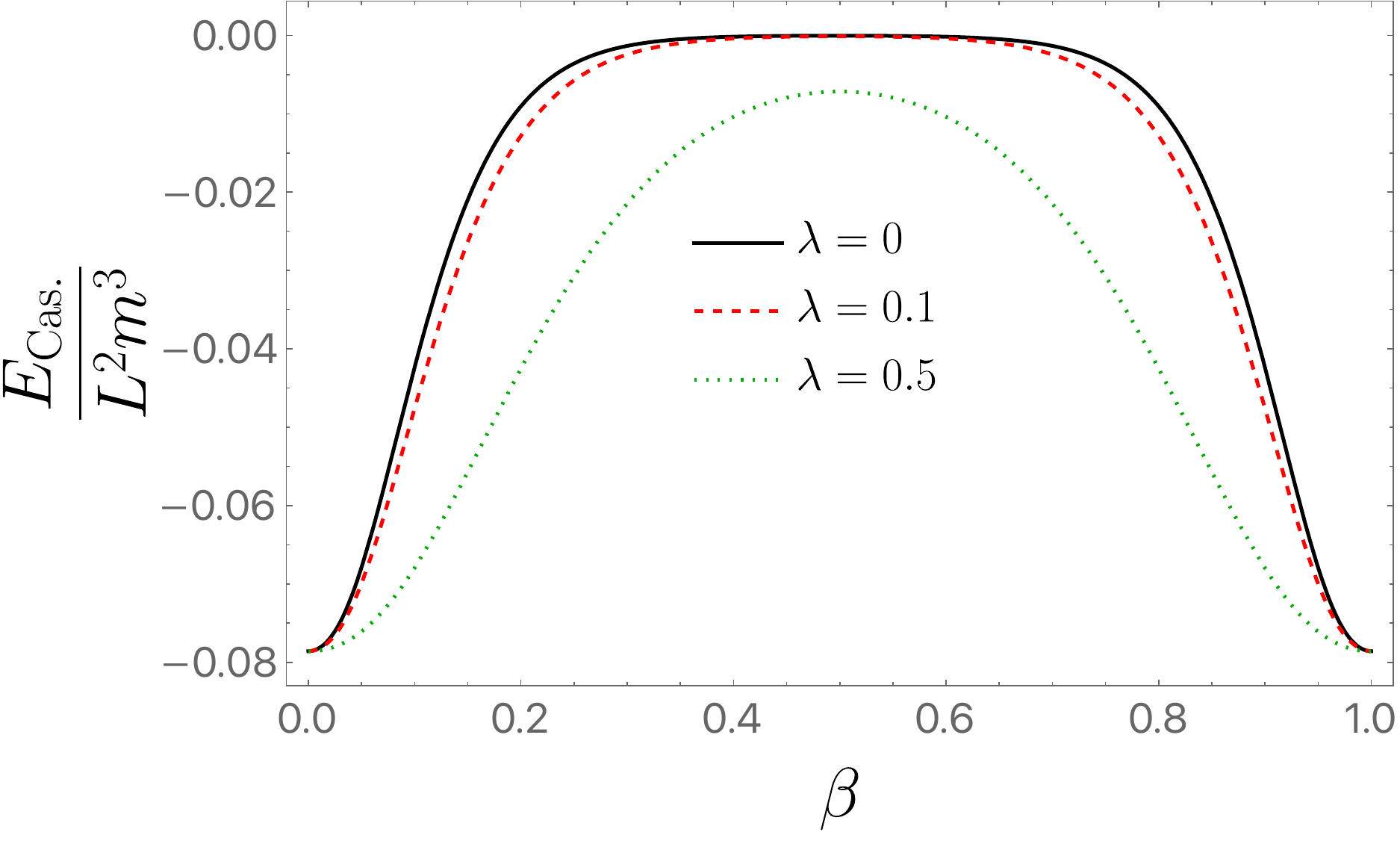}
\hfill
\includegraphics[width=.49\textwidth]{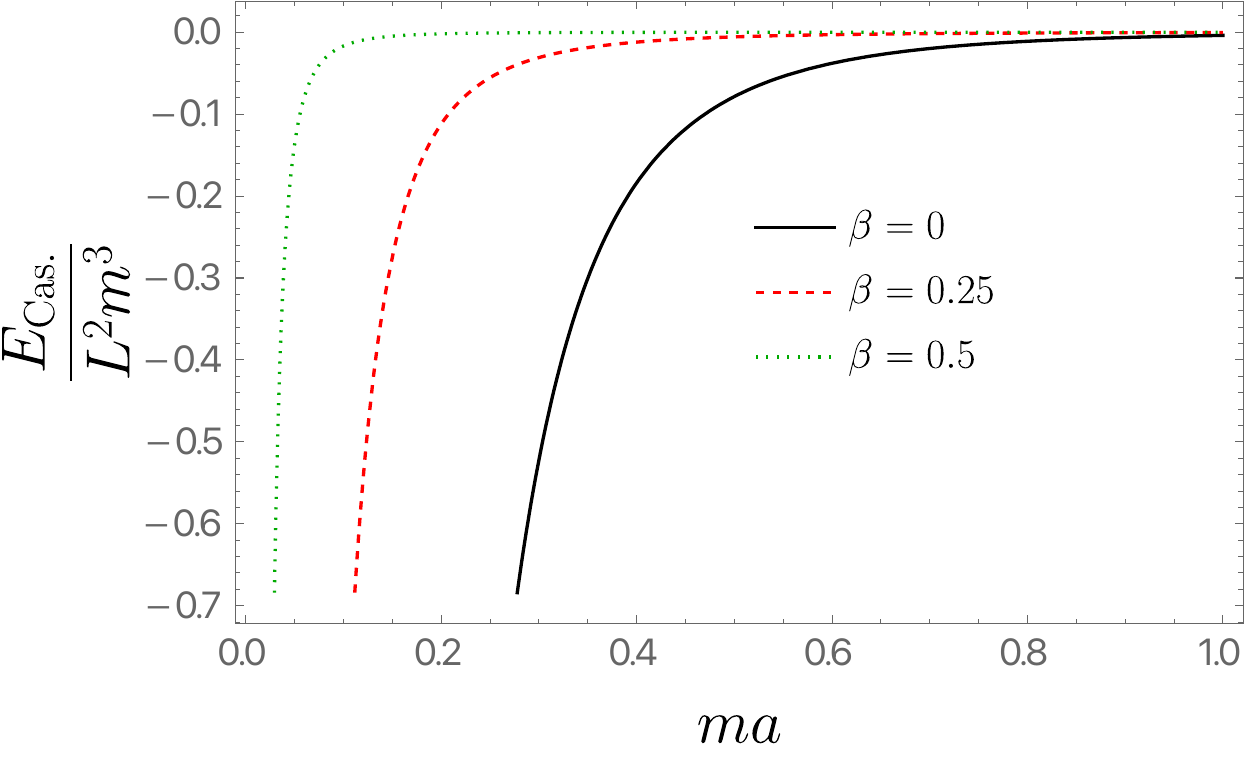}
\caption{ \label{ECasalphax5}
The left panel shows the scaled Casimir energy  ${E}_{\rm Cas.}/(L^2m^3)$ as a function of parameter $\beta$ for three values of Lorentz violation's intensity $\lambda=0, 0.1,0.5$ with fixed $ma=0.5$ and $q/a=0.5$ whereas the right panel shows the scaled Casimir energy as a function of $ma$ for various values of  $\beta=0,0.25, 0.5$ with fixed $q/a=0.5$ and $\lambda=0.1$. This figure presents the Lorentz violation in the $x^5$-direction.}
\end{figure} 
 
In Fig.~\ref{ECas3Dalphax5}, we show the behavior of the scaled Casimir energy  ${E}_{\rm Cas.}/(L^2m^3)$ as a function of $ma$ and $q/a$ for two various values of $\beta$ with a fixed value of the Lorentz violation parameter $\lambda=0.1$. The left panel is for $\beta=0$ whereas the right is for $\beta=0.5$. This figure shows the same behavior as in Fig.~\ref{ECas3Dalpha}. Namely, as the parameter $ma$ increases, the Casimir energy approaches zero and the one with $\beta=0.5$ more rapidly approaches zero compared to $\beta=0$. One can see this behavior more clearly in Figs.~\ref{ECasmax5} and \ref{ECasqax5}. Another interesting behavior is when we alter the parameter $\lambda$. As it increases, the magnitude of the Casimir energy increases. We note that this behavior is opposite to that of a space-like vector case for the $x^3$-direction. In Fig.~\ref{ECasqax5}, it can be inferred that the magnitude of the Casimir energy increases with a decrease in the parameter $q/a$ of both panels. In addition, the Casimir energy is less sensitive to the changing of the parameter $\lambda$. The magnitude of the Casimir energy is symmetric under the changing of parameter $\beta$, where the maximum amplitude of the Casimir energy for fixed $\lambda$ is achieved when $\beta=0$ whereas the minimum one is achieved when $\beta=0.5$ (Fig.~\ref{ECasalphax5}). This feature is consistent with that of Ref.~\cite{deFarias:2023xjf}, which discussed for the case of the EM field. It can also be observed that different $\lambda$ gives different minimum value of the Casimir energy.

Below, we discuss some limits of the Casimir energy in the case of Lorentz violations in $x^5$-directions. For the massless case, we have 
\begin{eqnarray}
E_{\rm Cas.}=-{L^2\over \pi^2}\sum_{\ell=-\infty}^\infty \int_{(1-\lambda)k_\ell}^\infty dz z (z^2-(1-\lambda)^2k^2_\ell)^{1/2} \ln\bigg(1+e^{-2 a z}\bigg). \label{formulax5Ecas}
\end{eqnarray}
By using the similar formula given in Eq.~\eqref{FormulaInt}, the integral part of the Casimir energy \eqref{casx5} with limit $q \gg a$ can be expressed as follows,
\begin{eqnarray}
\label{FormulaIntx5}
\int d{\ell}\int_{\sqrt{(1-\lambda)^2k^2_{\ell}+m^2}} dz (z^2-(1-\lambda)^2k^2_{\ell}-m^2)^{1/2}f(z)= {q\over 4 (1-\lambda)}
\int^\infty_m dz (z^2-m^2)f(z),
\end{eqnarray}
where we have used $d{\ell}=(q /2\pi (1-\lambda)) d((1-\lambda)k_\ell)$,
so that the Casimir energy in the case of $q \gg a$ reads 
\begin{eqnarray}
E_{\rm Cas.}=-{q L^2\over 4\pi^2 (1-\lambda)}\int^\infty_m dzz(z^2-m^2)  \ln\bigg(1+{z-m\over z+m}e^{-2 a z}\bigg). 
\end{eqnarray}
In the massless case, the above expression reduces to the following form
\begin{eqnarray}
E_{\rm Cas.}=-{45 qL^2 \zeta(5)\over 512\pi^2 (1-\lambda) a^4}. \label{Ecasm0x5}
\end{eqnarray}
Let us compare Eqs.\eqref{Ecasm0x3} and \eqref{Ecasm0x5}. We note that those two expressions differ because of how the factor $(1-\lambda)$ is treated in each case. The factor $(1-\lambda)a^4$ in Eq.~\eqref{Ecasm0x5} is treated as a single denominator and is not raised to any power. In contrast, in Eq.~\eqref{Ecasm0x3}, the factor $(1-\lambda)$ is explicitly raised to the fourth power and divided by $a^4$. It means that the factor $(1-\lambda)$ will amplify its effect depending on the value of $(1-\lambda)$. Considering their relative magnitude, when $(1-\lambda)<1$, the factor $(1-\lambda)^4$ will be much smaller than $(1-\lambda)$, making the latter expression is less sensitive than former due to the changes of $\lambda$.

We next consider the Casimir energy in the case of $q \ll a$ and $\beta=0$. In this case, we have
\begin{eqnarray}
E_{\rm Cas.}=-{L^2\over \pi^2}\int_m^\infty dz z (z^2-m^2)^{1/2} \ln\bigg(1+{z-m\over z+m}e^{-2 a z}\bigg). 
\end{eqnarray}
Introducing new variable $az=y+am$, the above expression becomes, 
\begin{eqnarray}
E_{\rm Cas.}=-{L^2\over \pi^2 a^3}\int_0^\infty dy (y+am) (y^2+2yam)^{1/2} \ln\bigg(1+{y\over y+2am}e^{-2 (y+am)}\bigg). 
\end{eqnarray}
One can see that in the case of $q\gg a$, the roles of Lorentz violation remain, where for $\lambda=0$, the result reduces to that of Ref.~\cite{Bellucci:2009hh}. Whereas, in the case of $q\ll a$ with $\beta=0$ as discussed above, the roles of Lorentz violation as well as compactified dimension disappear. In this case, the result coincides with that of Ref.~\cite{Bellucci:2009hh}, which discussed the preserve Lorentz symmetry. In the massless case together with $q \ll a$ with $\beta=0$, the above Casimir energy reduces to $E_{\rm Cas.}=7 \pi^2 L^2/2880a^3$.

\section{Casimir Pressure}
\label{CasimirPressure} 

In this section, we investigate the Casimir pressure by using the following formula
\begin{eqnarray}
P_{\rm Cas.}=-{1\over L^2}{\partial E_{\rm Cas.}\over \partial a}. 
\label{PCasformula}
\end{eqnarray}
Based on the previous result, we observe that the Casimir energy remains unaffected by time-like Lorentz violation. Equation~\eqref{PCasformula} shows that time-like Lorentz violations also do not impact the Casimir pressure. Therefore, we will focus our discussion on the Casimir pressure only in the cases of space-like vectors, specifically in both $x^3$- and $x^5$- directions.

\subsection{Space-like vector case in $x^3$-direction }
By using the Casimir energy $E_{\rm Cas.}$ given in Eq.~\eqref{ECas}, we have the Casimir pressure for the space-like vector case in $x^3$-directions as follows
\begin{eqnarray}
\label{Pcasx3}
    P_{\rm Cas.}=-{2\over (1-\lambda)\pi^2}\sum_{\ell=-\infty}^\infty\int_{m_\ell}^\infty dz {z^2 (z^2-m^2_\ell)^{1/2}\over 1+{z+m\over z-m}e^{2 b z}}. 
\end{eqnarray}
By introducing a new variable, 
\begin{equation}
    bz = x +bm_\ell \label{bz}
\end{equation}
the Casimir pressure can be rewritten as follows
\begin{eqnarray}
    P_{\rm Cas.}=-{2\over (1-\lambda)\pi^2b^4}\sum_{\ell=-\infty}^\infty\int_0^\infty dz {(x+bm_\ell)^2 (x^2+2xbm_\ell)^{1/2}\over 1+{x+bm_\ell+bm\over x+bm_\ell-bm}e^{2 (x+b m_\ell)}}.  
\end{eqnarray}

\begin{figure}[t!]
\centering 
\includegraphics[width=.49\textwidth]{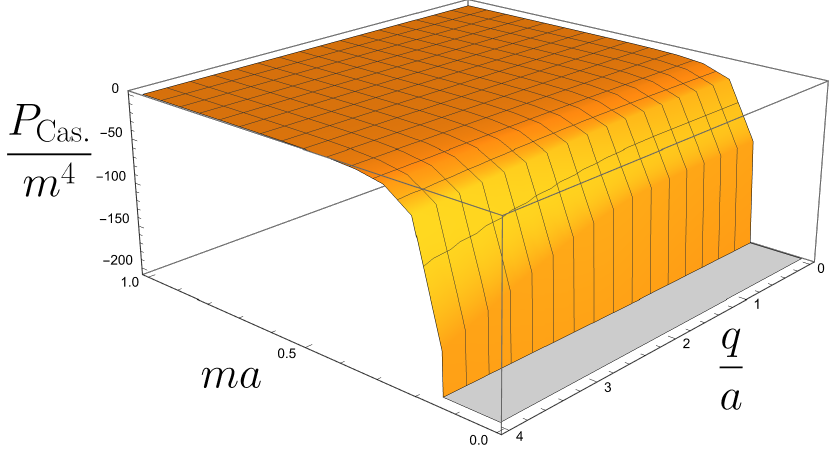}
\hfill
\includegraphics[width=.49\textwidth]{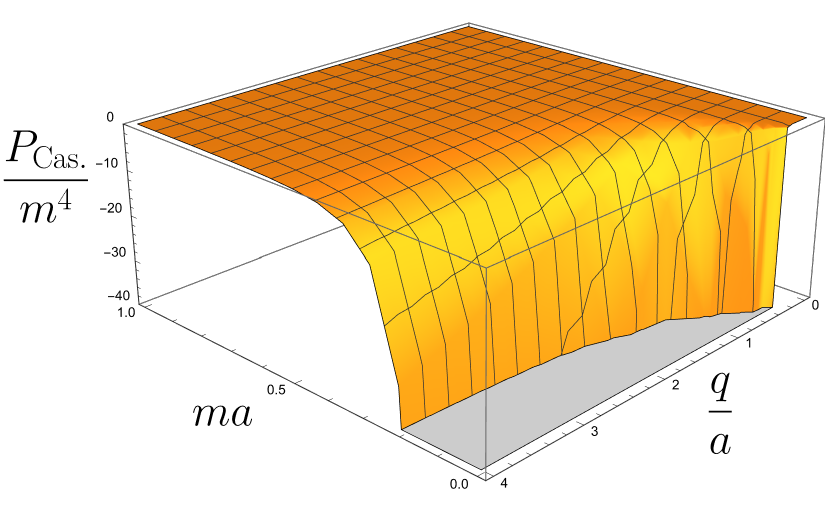}
\caption{\label{PCas3Dalpha} Plot of the scaled Casimir pressure $P_{\rm Cas.}/m^4$ as a function of $ma$ and $q/a$ with  fixed intensity of the Lorentz's violation $\lambda=0.1$ for two values of parameter $\beta$. For the left panel, we use $\beta=0$ while for the right panel, we use $\beta=0.5$. This figure presents the Lorentz violation in the $x^3$-direction.}
\end{figure}

\begin{figure}[t!]
\centering 
\includegraphics[height=.3\textwidth]{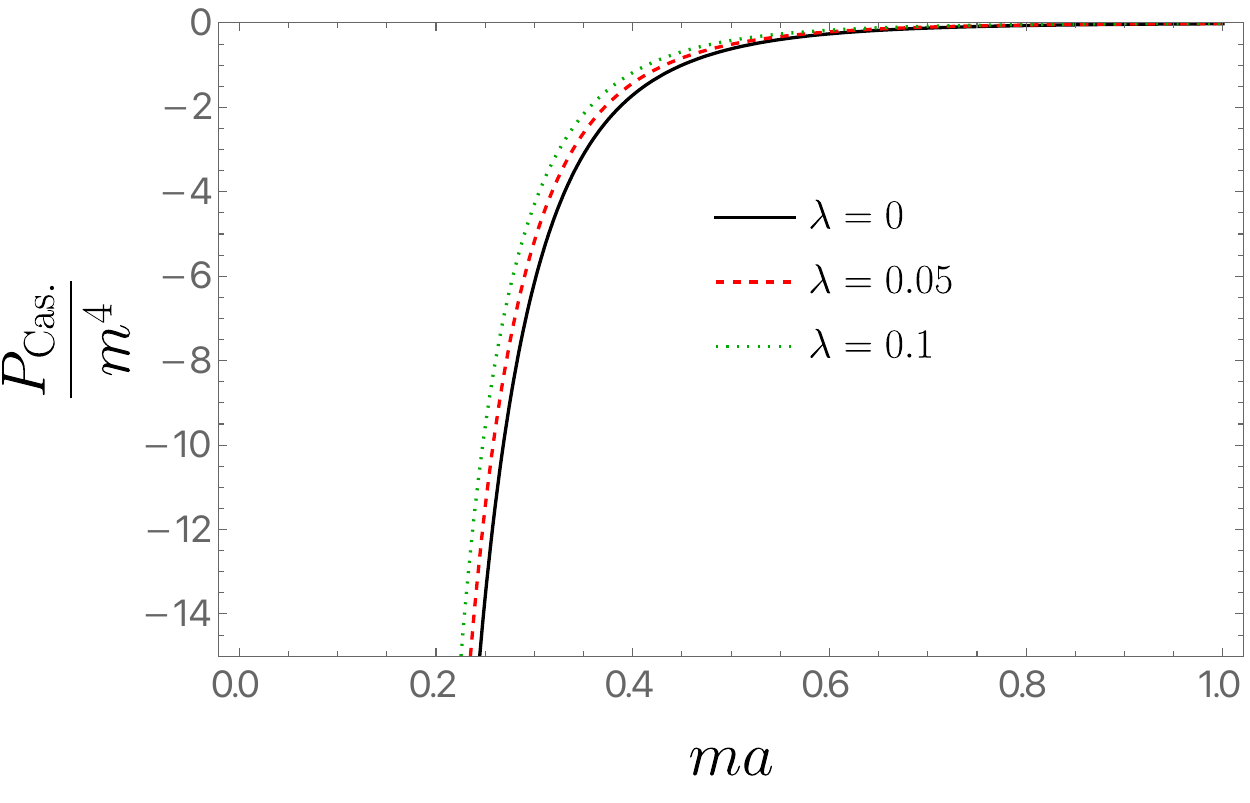}
\hfill
\includegraphics[height=.3\textwidth]{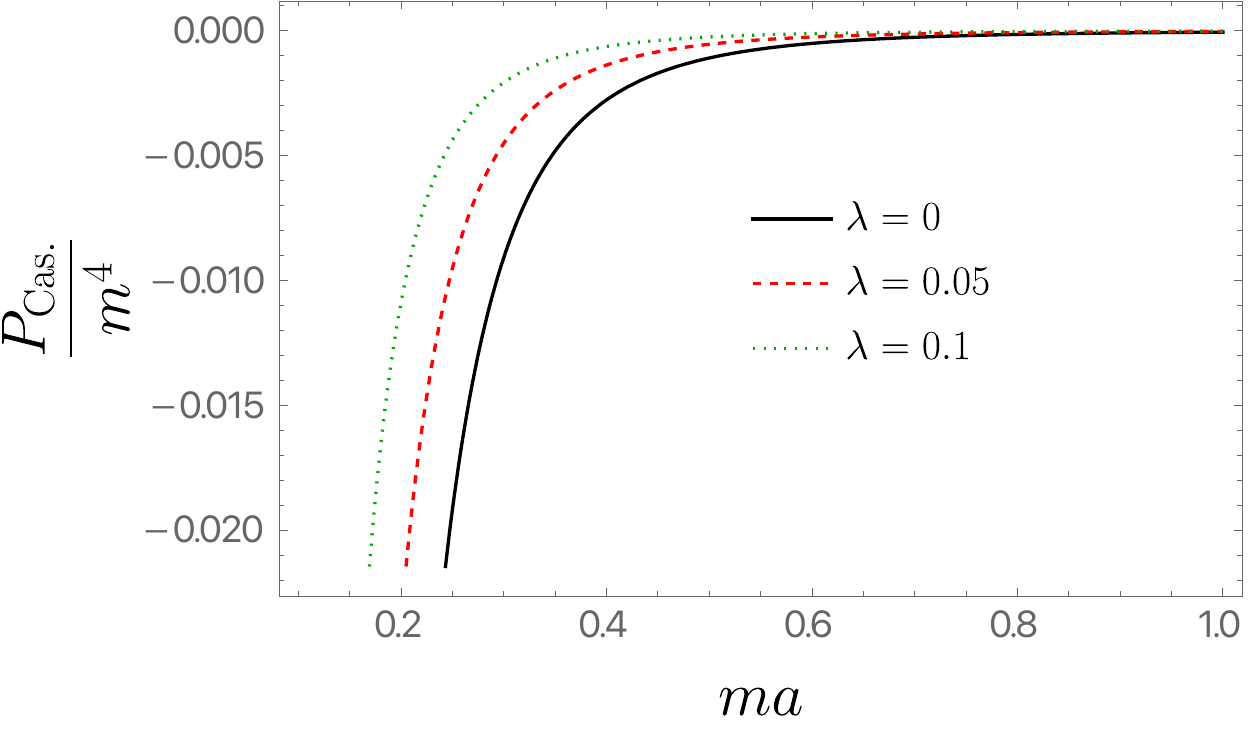}
\caption{\label{ECasma} Plot of the scaled Casimir pressure ${P}_{\rm Cas.}/m^4$ as a function of $ma$ for various value of the Lorentz violation's intensity $\lambda=0,0.05,0.1$ with fixed $q/a=0.5$ and two values of $\beta$. In the left panel, we use $\beta=0$ while in the right panel we use $\beta=0.5$. This figure shows the Lorentz violation in the $x^3$-direction.}  
\end{figure}

\begin{figure}[h!]
\centering 
\includegraphics[width=.49\textwidth]{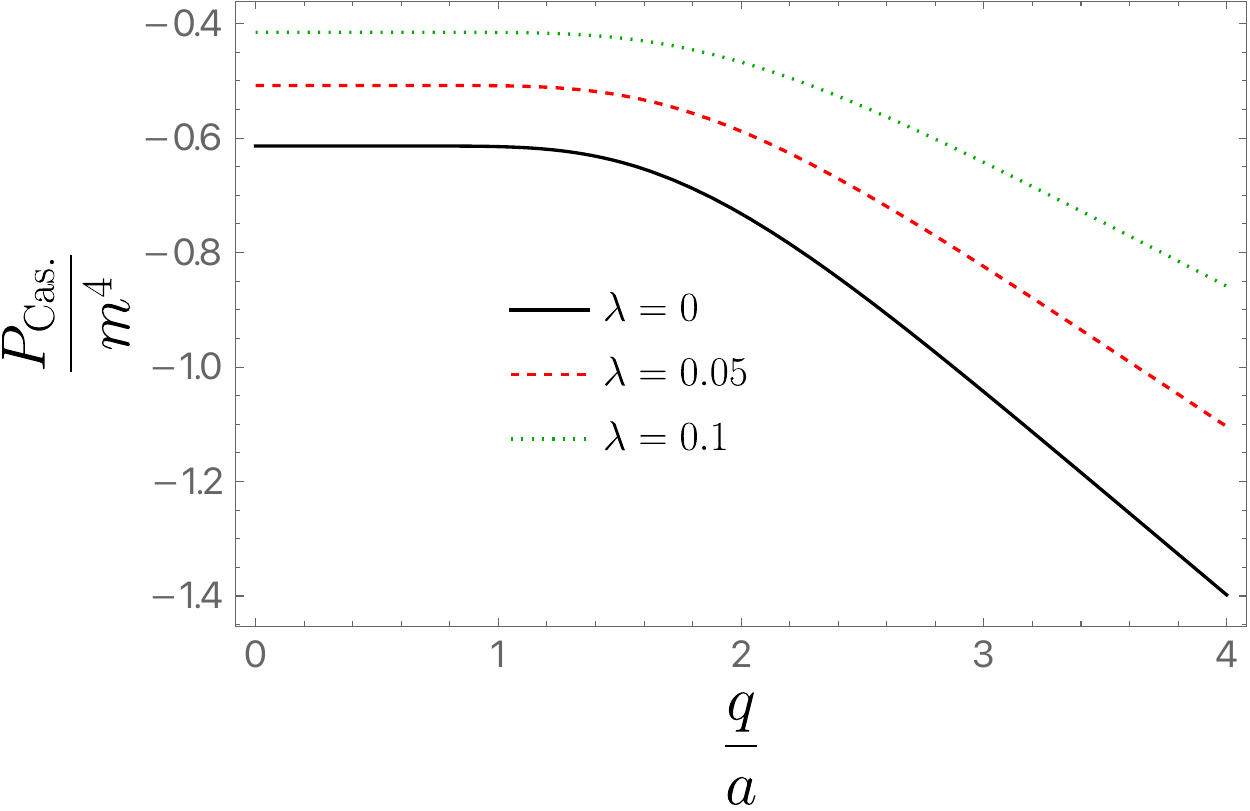}
\hfill
\includegraphics[width=.49\textwidth]{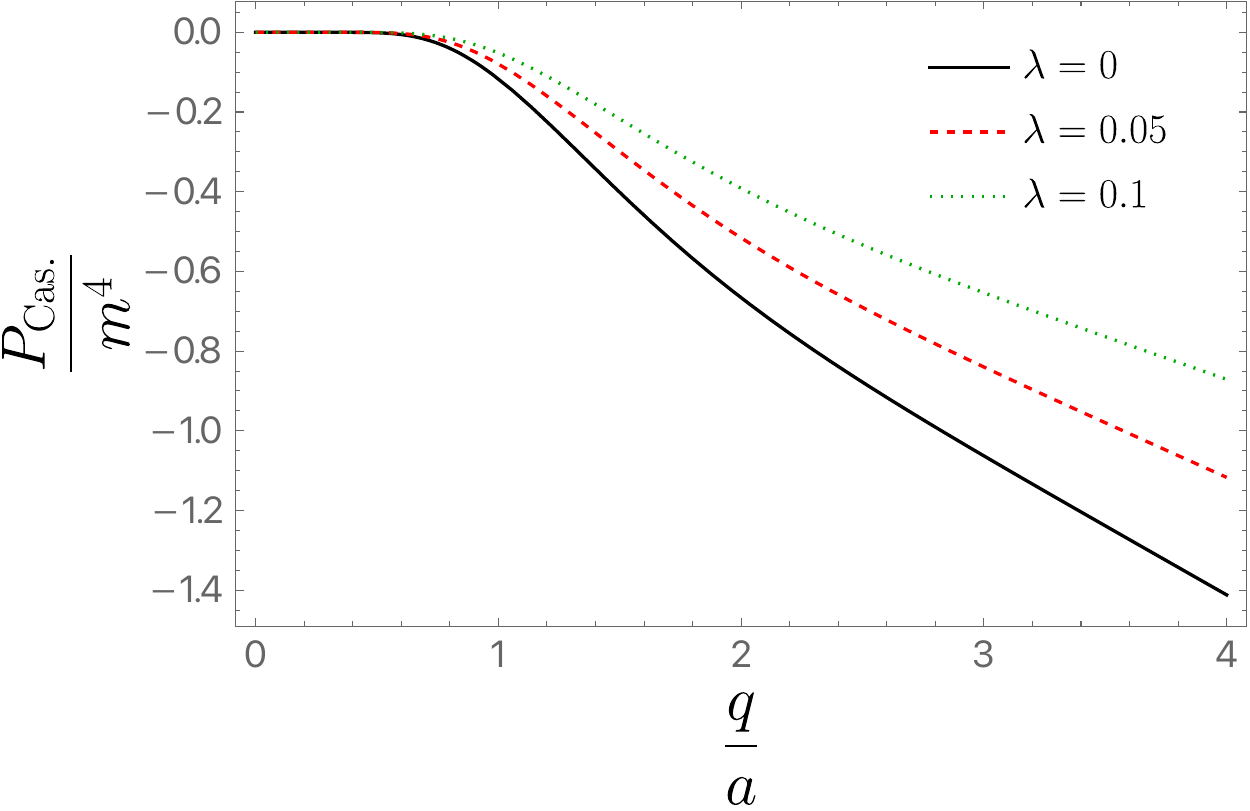}
\caption{\label{ECasqa} Plot of the scaled Casimir pressure ${P}_{\rm Cas.}/m^4$ as a function of $q/a$ for various values of the Lorentz violation's intensity $\lambda=0,0.05,0.1$ with fixed $ma=0.5$ and two values of $\beta$. In the left panel, we use $\beta=0$ while in the right panel, we use $\beta=0.5$. This figure presents the Lorentz violation in the $x^3$-direction.} 
\end{figure}

\begin{figure}[t!]
\centering 
\includegraphics[width=.49\textwidth]{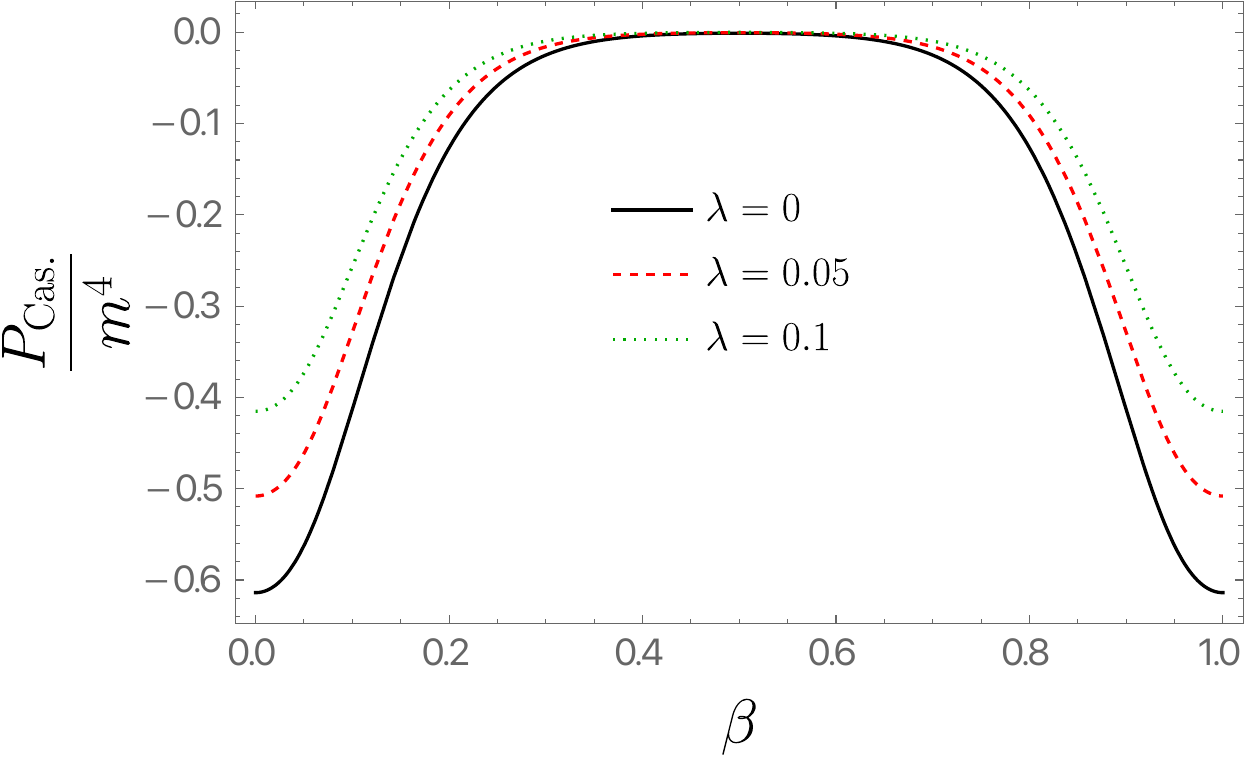}
\hfill
\includegraphics[width=.49\textwidth]{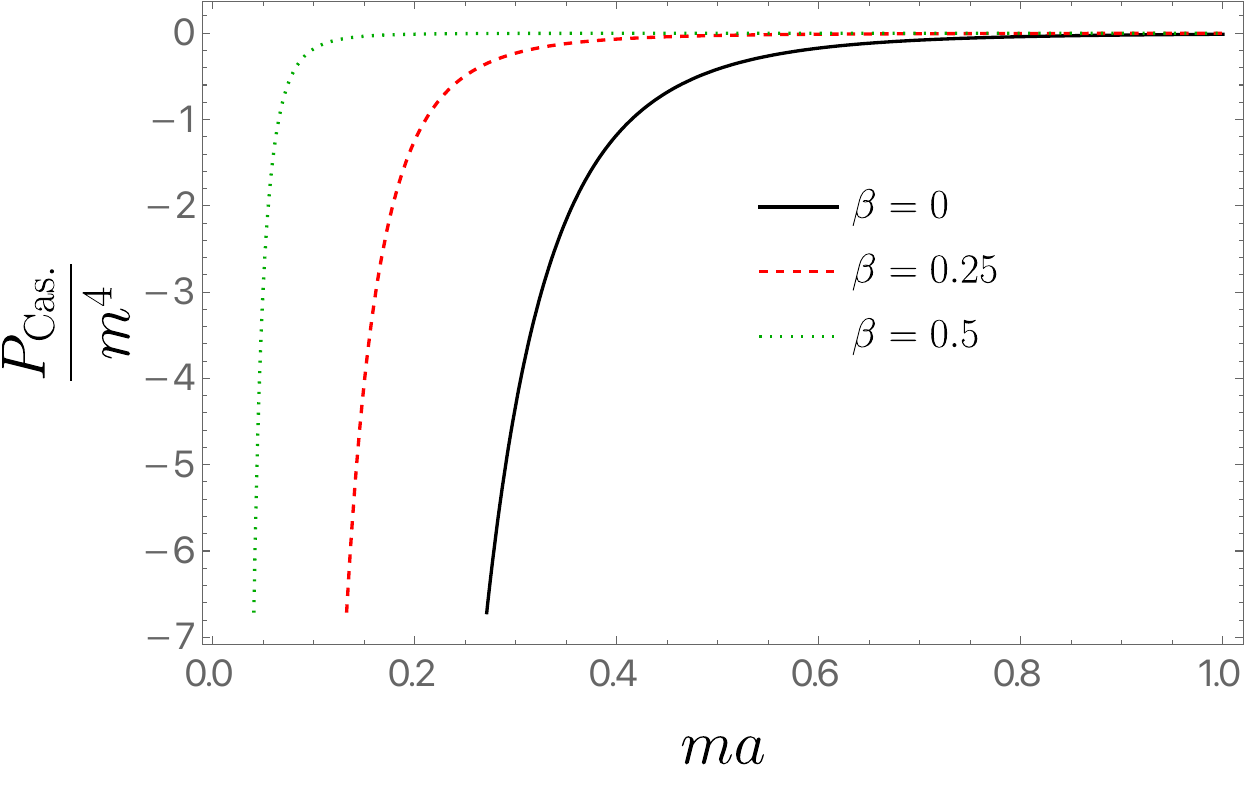}
\caption{ \label{PCasalpha}
The left panel shows the scaled Casimir pressure  ${P}_{\rm Cas.}/m^4$ as a function of parameter $\beta$ for three values of Lorentz violation's intensity $\lambda=0, 0.05,0.1$ with $ma=0.5$ and $q/a=0.5$ while the right panel shows the scaled Casimir pressure as a function of $ma$ with $q/a=0.5$, $\lambda=0.1$ and $\beta=0,0.25, 0.5$. This figure presents the Lorentz violation in the $x^3$-direction.}
\end{figure}

We numerically study the behavior of the Casimir pressure in Figs. \ref{PCas3Dalpha}, \ref{ECasma}, \ref{ECasqa}, and \ref{PCasalpha}. In general, the behavior of the Casimir pressure is similar to that of Casimir energy. In Fig.~\ref{PCas3Dalpha}, we show the behavior of the scaled Casimir pressure as a function of $ma$ and $q/a$ for two various values of $\beta$ with fixed value of the Lorentz violation $\lambda=0.1$. The left panel is for $\beta=0$ while the right one is for $\beta=0.5$. From this figure, one can see that the Casimir energy goes to zero as the parameter $ma$ increases, which is confirmed by Fig.~\ref{ECasma}. The magnitude of the Casimir energy is symmetric with respect to the parameter $\beta$, where the  maximum is achieved when $\beta=0$ while the minimum is achieved when $\beta=0.5$ (see left panel of Fig.~\ref{PCasalpha}).

In the following, we discuss the Casimir pressure in some limits. Taking massless limit from Eq.~\eqref{Pcasx3}, the Casimir pressure reads
\begin{eqnarray}
   P_{\rm Cas.}=-{2\over (1-\lambda)\pi^2}\sum_{\ell=-\infty}^\infty\int_{k_\ell}^\infty dz {z^2 (z^2-k^2_\ell)^{1/2}\over 1+e^{2 b z}}.
\end{eqnarray}
For the case of $q\gg a$, treating it with the same procedure as in Eqs.~\eqref{casen} and \eqref{FormulaIntx3}, the Casimir pressure \eqref{Pcasx3} becomes,
\begin{eqnarray}
 P_{\rm Cas.}= -{q\over 2 (1-\lambda)\pi^2}\int_{m}^\infty dz {z^2 (z^2-m^2)\over 1+{z+m\over z-m}e^{2 b z}}.
\end{eqnarray}
Next, we consider the Casimir energy in the case of $q \ll a$ and $\beta=0$. In this case we have
\begin{eqnarray}
    P_{\rm Cas.}=-{2\over (1-\lambda)\pi^2}\int_{m}^\infty dz {z^2 (z^2-m^2)^{1/2}\over 1+{z+m\over z-m}e^{2 b z}}. 
\end{eqnarray}
In the massless case together with $q \ll a$ and $\beta=0$, the Casimir pressure reduces to $P_{\rm Cas.}=-7 \pi^2/960 b^4$.  

\subsection{Space-like vector case in $x^5$-direction }

From the Casimir energy $E_{\rm Cas.}$ in Eq.~\eqref{casx5}, through formula in Eq.\eqref{PCasformula}, we can obtain the Casimir pressure for the space-like vector case in $x^5$-directions as follows
\begin{eqnarray}
    P_{\rm Cas.}=-{2\over \pi^2}\sum_{\ell=-\infty}^\infty\int_{\tilde m_\ell}^\infty dz {z^2 (z^2-\tilde m^2_\ell)^{1/2}\over 1+{z+m\over z-m}e^{2 a z}}. \label{pcasx5}
\end{eqnarray}
By introducing a new variable, 
\begin{equation}
    az = x +a\tilde m_\ell, 
    \label{az}
\end{equation}
the above Casimir pressure can be rewritten as follows
\begin{eqnarray}
    P_{\rm Cas.}=-{2\over \pi^2a^4}\sum_{\ell=-\infty}^\infty\int_0^\infty dx {(x+a\tilde m_\ell)^2 (x^2+2xa \tilde m_\ell)^{1/2}\over 1+{x+a \tilde m_\ell+am\over x+a \tilde m_\ell-am}e^{2 (x+a \tilde m_\ell)}}.  
\end{eqnarray}

\begin{figure}[tbp]
\centering 
\includegraphics[width=.49\textwidth]{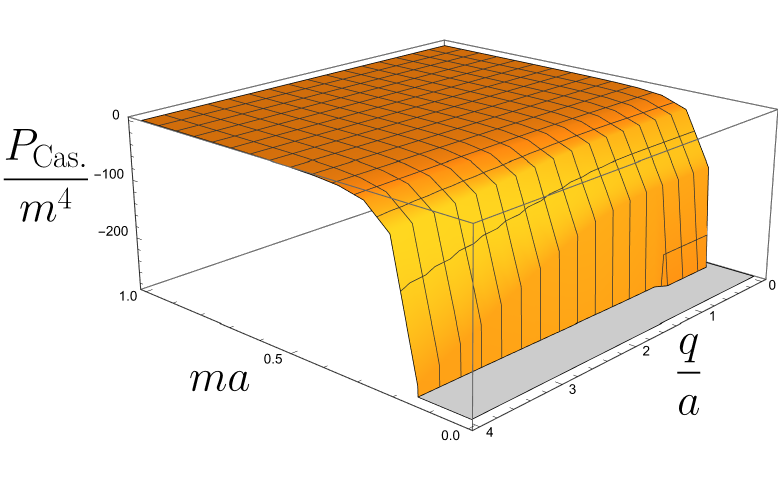}
\hfill
\includegraphics[width=.49\textwidth]{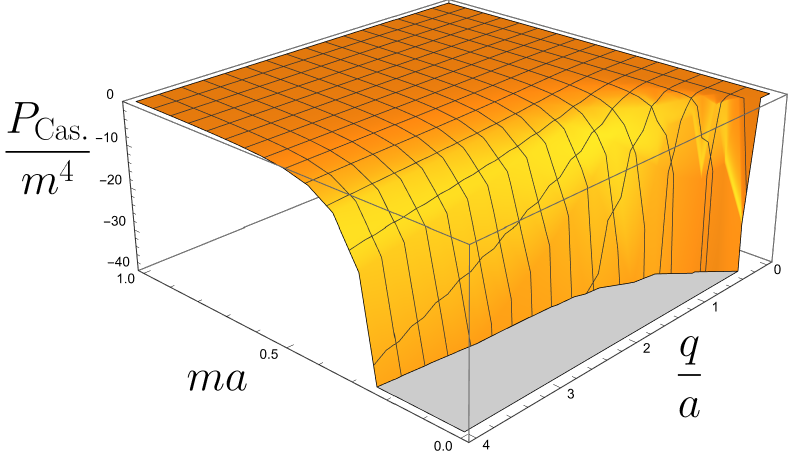}
\caption{\label{PCas3Dalphax5} Plot of the scaled Casimir pressure ${P}_{\rm Cas.}/m^4$ as a function of $ma$ and $q/a$  for two values of parameter $\beta$ with fixed value of Lorentz violation intensity $\lambda=0.1$. In the left panel, we use $\beta=0$ while in the right panel we use $\beta=0.5$. This figure presents the Lorentz violation in the $x^5$-direction.}
\end{figure}

\begin{figure}[tbp]
\centering 
\includegraphics[width=.49\textwidth]{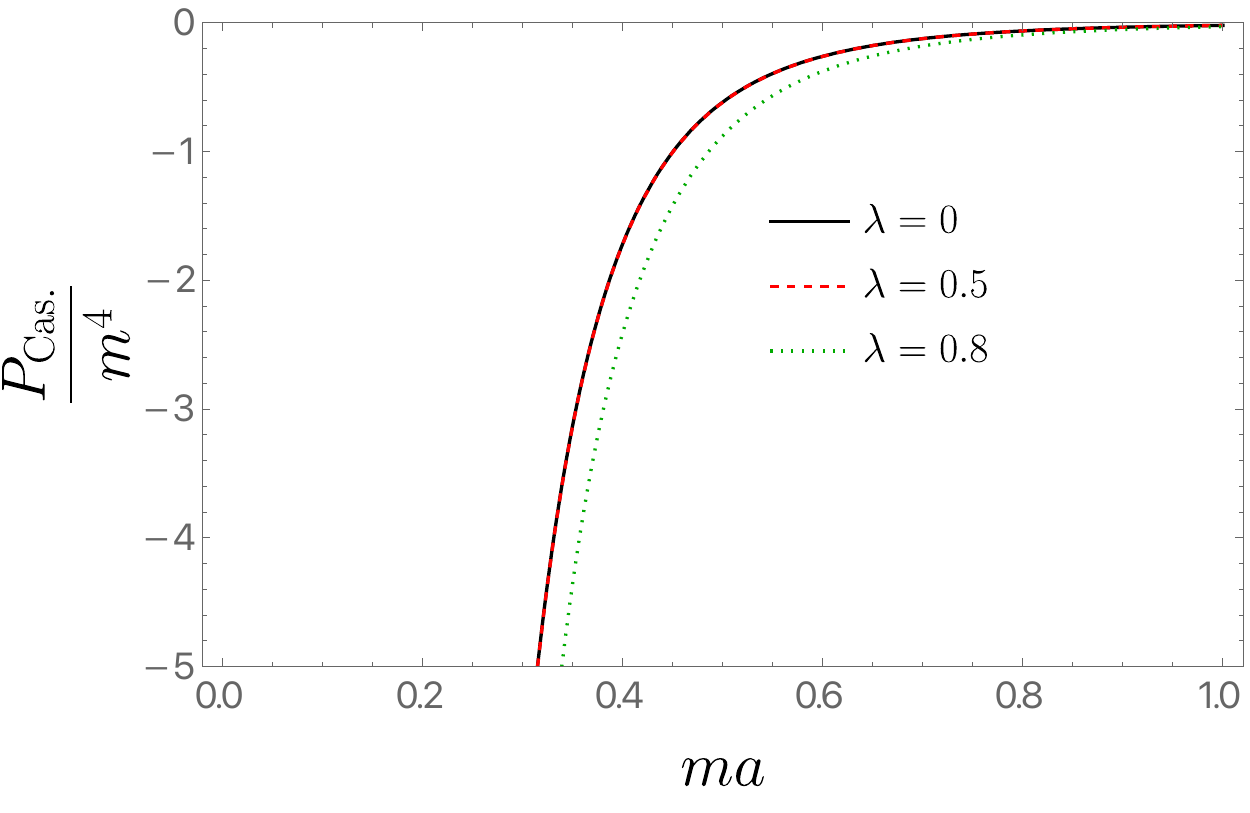}
\hfill
\includegraphics[width=.49\textwidth]{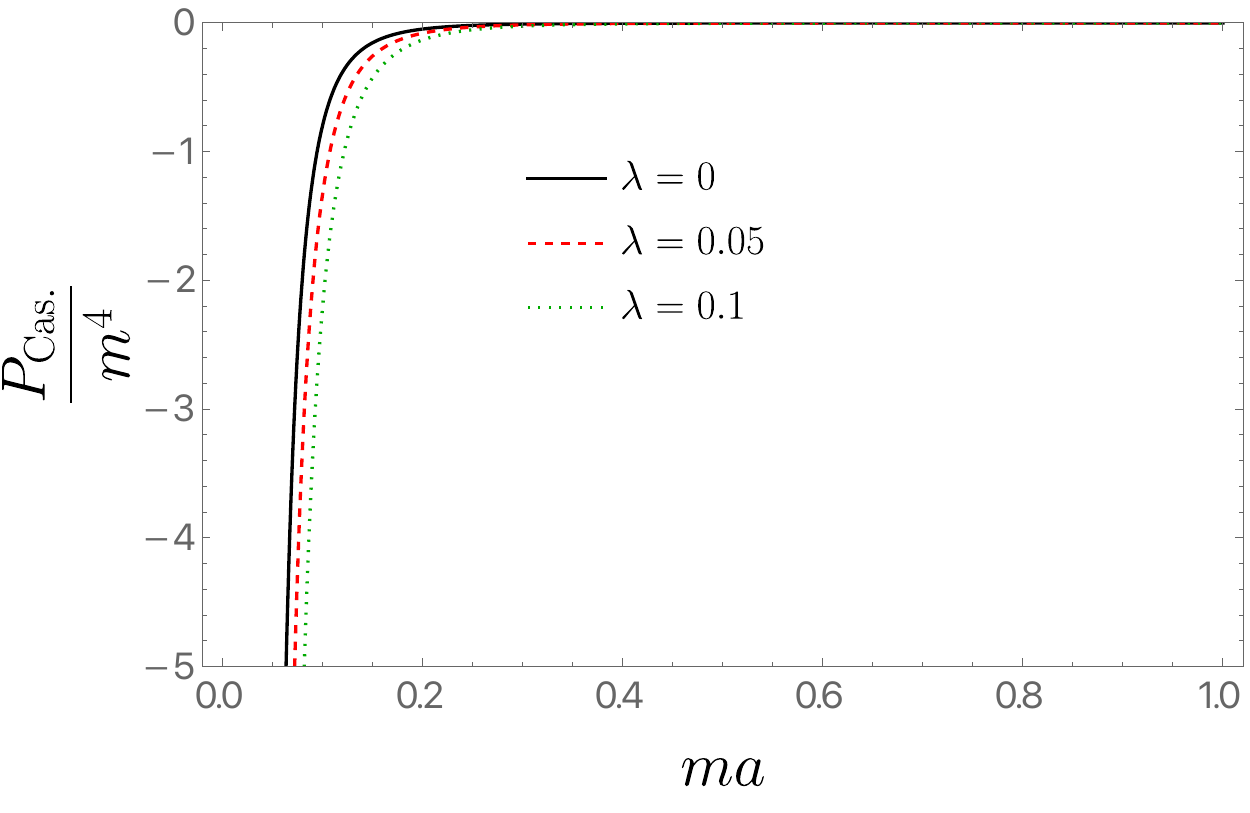}
\caption{\label{PCasmax5} 
Plot of the scaled Casimir pressure ${P}_{\rm Cas.}/m^4$ as a function of $ma$ for various value of the Lorentz violation's intensity $\lambda$ with fixed $q/a=0.5$ and two values of $\beta$. In the left panel, we use $\beta=0$ and $\lambda=0,0.5,0.8$  while in the right panel we use $\beta=0.5$ and $\lambda=0,0.05,0.1$. This figure presents the Lorentz violation in the $x^5$-direction. }
\end{figure}

\begin{figure}[tbp]
\centering 
\includegraphics[width=.49\textwidth]{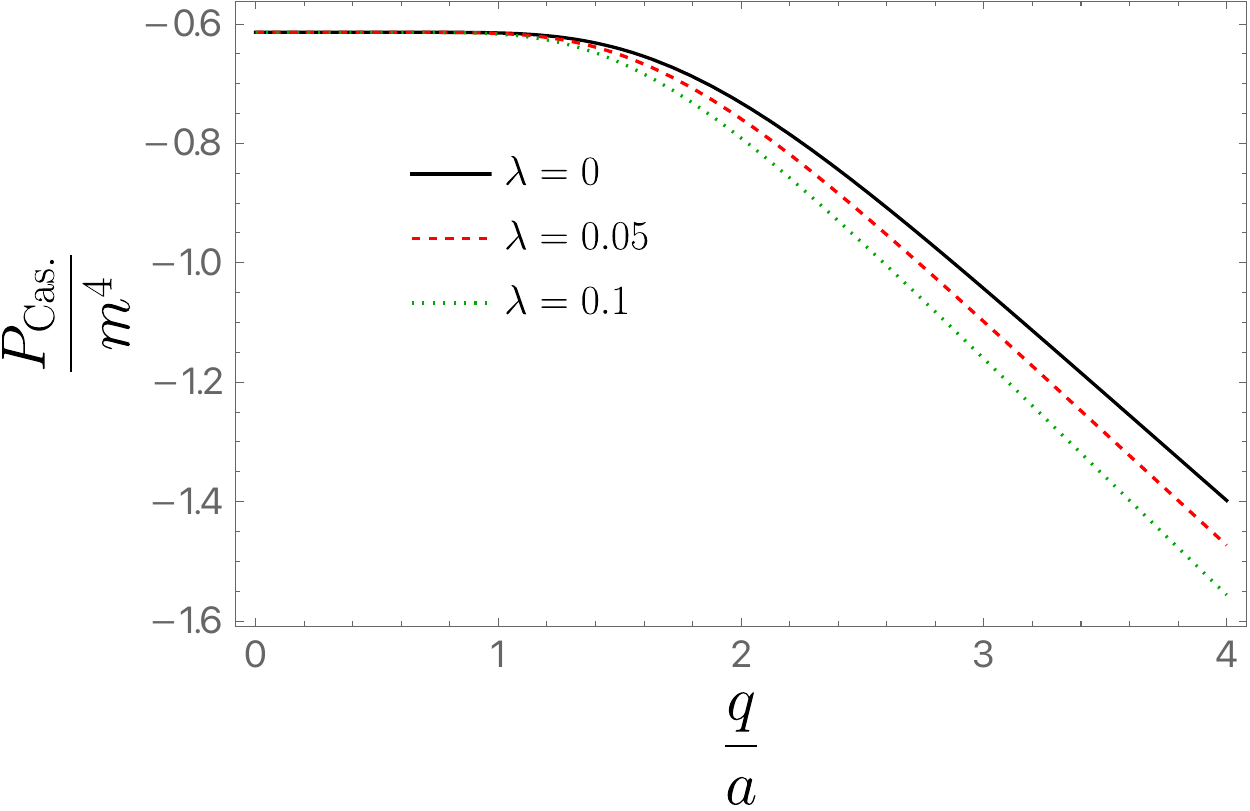}
\hfill
\includegraphics[width=.49\textwidth]{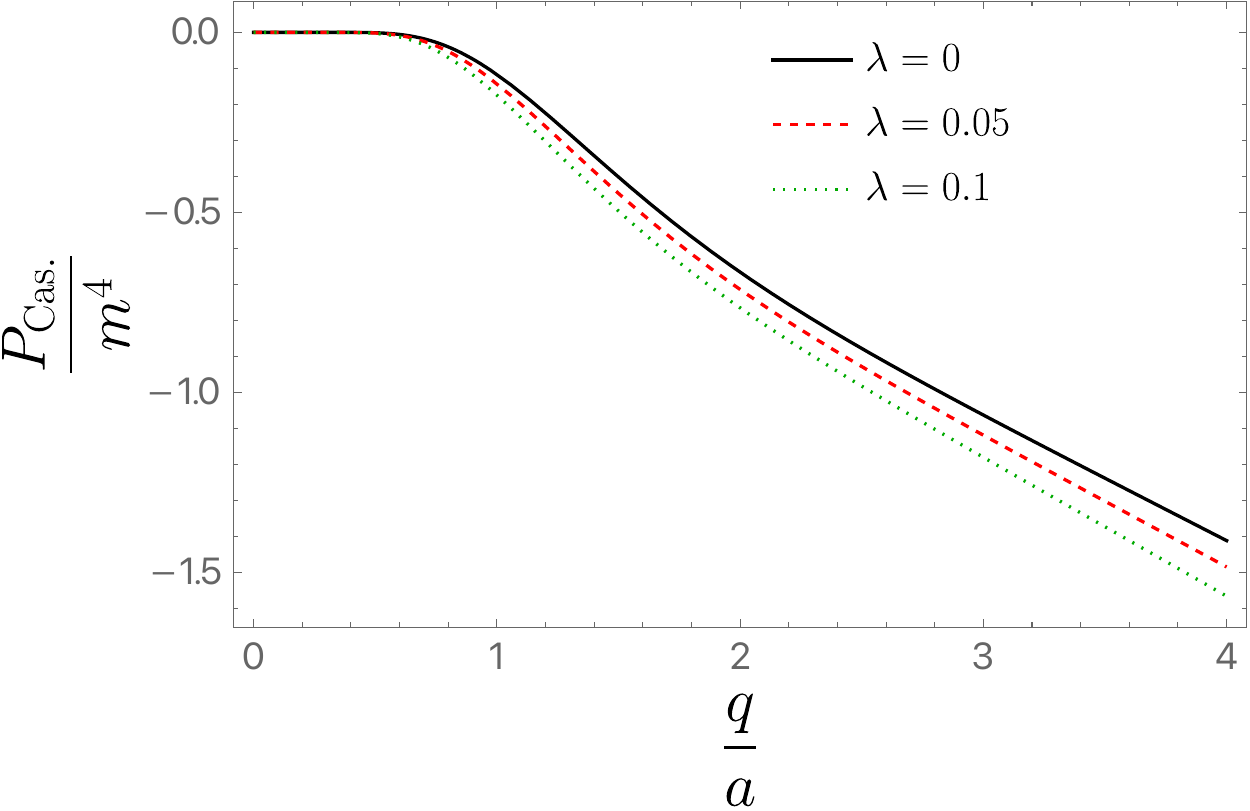}
\caption{\label{PCasqax5}Plot of the scaled Casimir pressure ${P}_{\rm Cas.}/m^4$ as a function of $q/a$ for various value of the Lorentz violation's intensity $\lambda=0,0.05,0.1$ with fixed $ma=0.5$ and two values of $\beta$. In the left panel, we use $\beta=0$ while in the right panel we use $\beta=0.5$. This figure presents the Lorentz violation in the $x^5$-direction.}
\end{figure} 

\begin{figure}[tbp]
\centering 
\includegraphics[width=.49\textwidth]{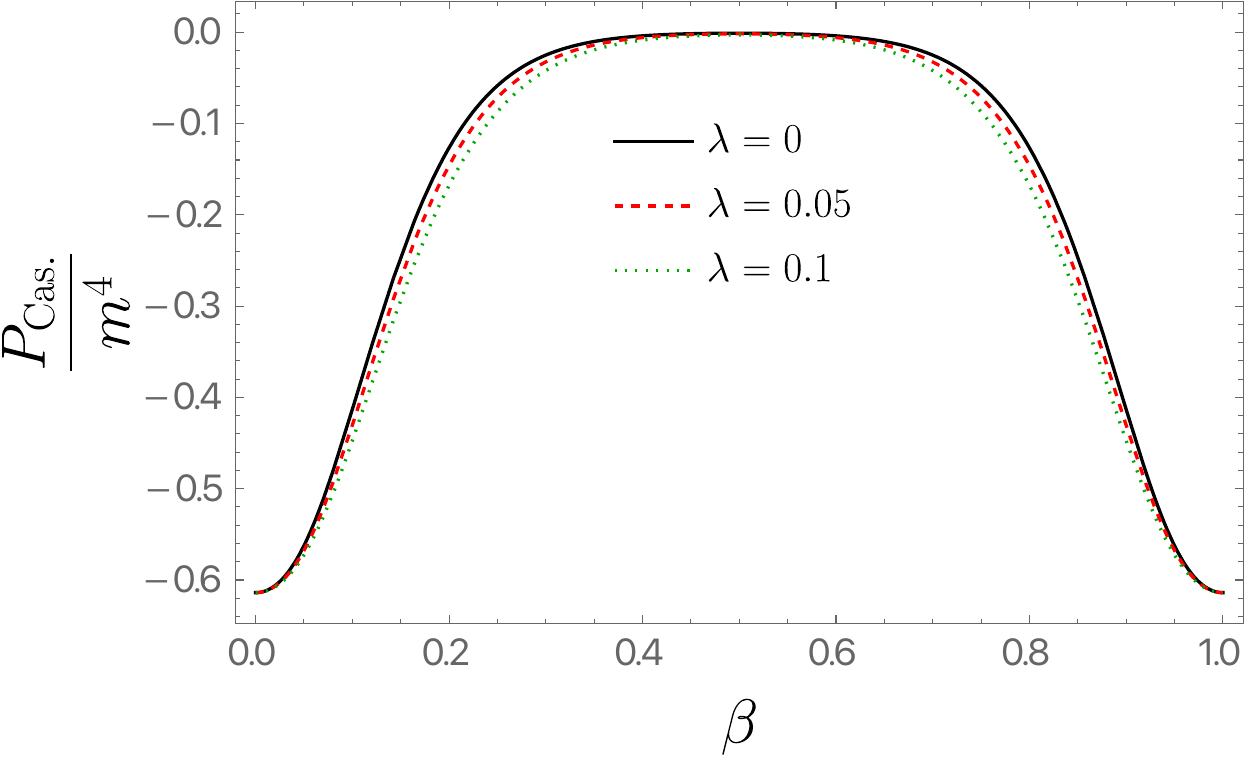}
\hfill
\includegraphics[width=.49\textwidth]{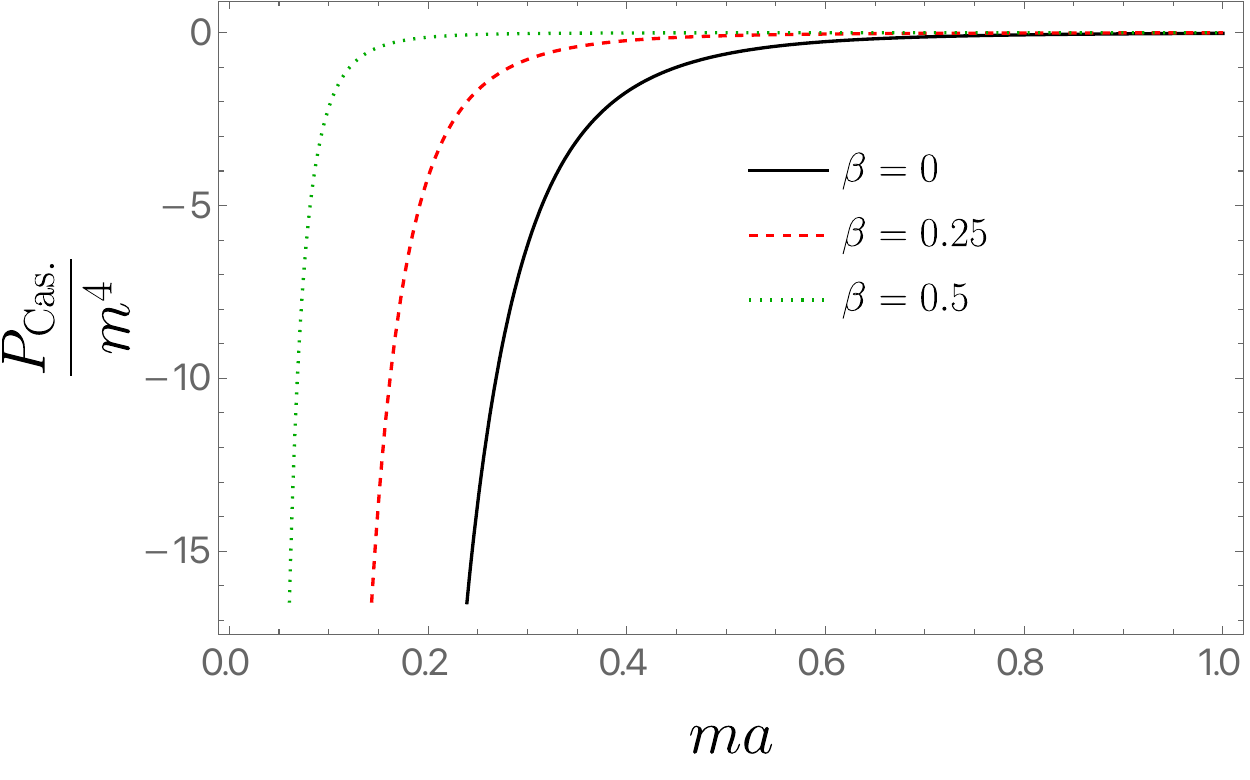}
\caption{ \label{PCasalphax5}
The scaled Casimir pressure ${P}_{\rm Cas.}/m^4$. The left panel shows the scaled Casimir pressure  ${P}_{\rm Cas.}/m^4$ as a function of parameter $\beta$ for three values of Lorentz violation's intensity $\lambda=0, 0.5,0.8$ with $\beta=0$  while the right panel shows the scaled Casimir energy as a function of $ma$ for various values of  $\beta=0,0.25, 0.5$ with fixed $q/a=0.5$ and $\lambda=0.1$.  This figure presents the Lorentz violation in the $x^5$-direction.}
\end{figure}

In Figs.~\ref{PCas3Dalphax5}-\ref{PCasalphax5}, we demonstrate the behavior of the scaled Casimir pressure, where the Lorentz violation is in $x^5$-direction. These figures are qualitatively similar to those of the Casimir energy. In Fig.~\ref{PCas3Dalphax5}, we plot the Casimir pressure as a function of the parameter $ma$ and 
 $q/a$ for fixed values of $\beta$ and $\lambda$. From these figures, one can see that the pressure goes to zero as the increase in the parameter $ma$ (see also Fig.~\ref{PCasmax5}). In Fig.~\ref{PCasqax5}, we plot the scaled Casimir pressure as a function of $q/a$ for various values of $\lambda$ with $\beta=0$ and $\beta=0.5$. From this figure, it can be inferred that the scaled the amplitude of the Casimir pressure increases as the $q/a$ increases.

As has been discussed in the Section \ref{CasimirEnergy} on Casimir energy, in what follows, we should also investigate the Casimir pressure for certain limits. Taking massless limit from Eq.~\eqref{pcasx5}, the Casimir pressure reads
\begin{eqnarray}
   P_{\rm Cas.}=-{2\over \pi^2}\sum_{\ell=-\infty}^\infty\int_{(1-\lambda)k_\ell}^\infty dz {z^2 (z^2-(1-\lambda)^2k^2_\ell)^{1/2}\over 1+e^{2 a z}}.
\end{eqnarray}
For the case that $q\gg a$, following procedure in Eqs.\eqref{formulax5Ecas} and \eqref{FormulaIntx5}, the Casimir pressure \eqref{pcasx5} is expressed as follows,
\begin{eqnarray}
 P_{\rm Cas.}= -{q\over 2 \pi^2 (1-\lambda)}\int_{m}^\infty dz {z^2 (z^2-m^2)\over 1+{z+m\over z-m}e^{2 a z}}. 
\end{eqnarray}
Next, we consider the Casimir energy in the case of $q \ll a$ with $\beta=0$  that results in 
\begin{eqnarray}
    P_{\rm Cas.}=-{2\over \pi^2}\int_{m}^\infty dz {z^2 (z^2-m^2)^{1/2}\over 1+{z+m\over z-m}e^{2 a z}}. 
\end{eqnarray}
In the massless case together with $q \ll a$ with $\beta=0$, the Casimir pressure reduces to $P_{\rm Cas.}=-7 \pi^2/960 a^4$.

\section{Frequency shift and size of the extra dimension}
\label{ExtraDim}

In this section, we study the phenomenological aspect of the Casimir force due to the existence of the extra dimension. We can study it via a resonant squared frequency shift as discussed in Refs.~\cite{deFarias:2023xjf, Pascoal:2008rna, Bressi:2000iw}. The size of the extra dimension is a parameter that is accessible to the experiment instrument. However, the Casimir force in the context of fermion fields has not yet been experimentally observed so far. 
Therefore, we will discuss below how to probe this phenomenon using existing experimental data from the Casimir force of the EM field due to an influence of the extra dimension.

To facilitate this study, we follow the procedure outlined in Ref.~\cite{deFarias:2023xjf}. We begin by writing the shift of the resonator's frequency, denoted as $\Delta v^2$, under the effect of any distance-dependent force $F$ as follows \cite{Bressi:2000iw},
\begin{eqnarray}
\Delta v^2  = -{L^2\over 4\pi^2 m_{\rm eff}}{\partial  \over \partial a}\left[ \frac{F}{L^2} \right] =-{L^2\over 4\pi^2 m_{\rm eff}}{\partial P\over \partial a}, \label{deltav2}
\end{eqnarray}
where $m_{\rm eff}$ is the effective mass, which is related to the properties of the system. 
We can compute the frequency shift of a massless fermionic field with Lorentz violation in  $x^3$- and $x^5$-directions using the result in the previous section. 
For the frequency shift with Lorentz violation in the $x^3$-direction, we have 
\begin{align}
    (\Delta v^2)_{x^{3}}=-{\hbar c (1-\lambda)^3 \over a^{5}\pi^4 } \left(\frac{L^2}{m_{\rm eff}}\right)_{F} \sum_{\ell=-\infty}^\infty\int_0^\infty dx {\left(x+\frac{a}{1-\lambda} k_\ell\right)^3 \left(x^2+2x \frac{a}{1-\lambda} k_\ell\right)^{1/2} e^{2 \left(x+\frac{a}{1-\lambda} k_\ell\right)}  \over \left[ 1+ e^{2 \left(x+\frac{a}{1-\lambda} k_\ell\right)}\right]^2},  \label{deltax3}
\end{align}
whereas the frequency shift with Lorentz violation in $x^5$-direction is given by
\begin{align}
    (\Delta v^2)_{x^{5}}=-{\hbar c \over a^{5}\pi^4 } \left(\frac{L^2}{m_{\rm eff}}\right)_{F} \sum_{\ell=-\infty}^\infty\int_0^\infty dx {(x+a(1-\lambda)k_\ell)^3 (x^2+2xa(1-\lambda)k_\ell)^{1/2} e^{2 (x+a (1-\lambda)k_\ell)}  \over ( 1+ e^{2 (x+a (1-\lambda)k_\ell)})^2}
    \label{deltax5}
\end{align}
Note that, in the above expression, we have included the constants $\hbar$ and $c$, which are useful when comparing the model with the experimental data.

As mentioned above, to obtain the result of the estimation for the size of the extra dimension, we will use a scale value of the experimental data from the EM field. Recall that the Casimir force for EM field is given by \cite{Casimir:1948dh} 
\begin{eqnarray}
    F_{\rm EM}=-{\hbar c\pi^2L^2\over 240 a^4},
\end{eqnarray}
so that the frequency shift is given by 
\begin{eqnarray}
\Delta v^2_{\rm EM}=-{\hbar c \over 240 a^5} \bigg( {L^2\over m_{\rm eff}}\bigg)_{\rm EM}
\label{fqshiftEM}
\end{eqnarray}
where the value of $(L^2/m_{\rm eff})_{\rm{EM}}=1.746~{\rm Hz^2m^3N^{-1}}$\cite{Bressi:2000iw} as has been suggested in Ref.~\cite{deFarias:2023xjf}. Since it is the only information we know, one may define the following ratio related to the fermionic case,
\begin{eqnarray}
    \left(\frac{L^2}{m_{\rm eff}}\right)_{F} := \alpha \left(\frac{L^2}{m_{\rm eff}}\right)_{\rm{EM}}, \label{alpha}
\end{eqnarray}
in which $\alpha$ is the constant quantity that determines how large the difference between two cases and its value can be larger, smaller, or equal to one. 

Next, the Casimir force of the fermionic field in the massless case under standard field theory is computed as follows,
\begin{eqnarray}
  F=-{7 \hbar c  \pi^2 L^2 \over 960 a^4},
\end{eqnarray}
where the frequency shift $\Delta v^2_{F}$ is given by
\begin{eqnarray}
\Delta v^2_F=-{7\hbar c  \over 960 a^5 } \left(\frac{L^2}{m_{\rm eff}}\right)_{F} = {7\alpha\over 4} \Delta v^2_{\rm EM},
\label{fqshiftFer}
\end{eqnarray}
where we have used Eqs. \eqref{fqshiftEM} and \eqref{alpha}. The experimental data show that $\Delta v^2_{\rm EM}=-C^{\rm EM}_{\rm Cas}/a^5$, where $C^{\rm EM}_{\rm Cas.}=(2.34\pm 0.34)\times 10^{-28}~{\rm Hz^2 m^5}$ \cite{Bressi:2002fr}. 
With all of this procedure, we can set Eq.~\eqref{fqshiftFer} as the experimental reference when performing numerical calculations. Note that when we carry out the numerical simulation for Eqs.~\eqref{deltax3} and \eqref{deltax5}, we should also take into account Eq.~\eqref{alpha}. Additionally, we can investigate the parameter values of $\lambda$ and $\beta$ that will be close to our experimental reference with a certain value of $\alpha$. 

\begin{figure}[tbp]
\centering 
\includegraphics[width=.65\textwidth]{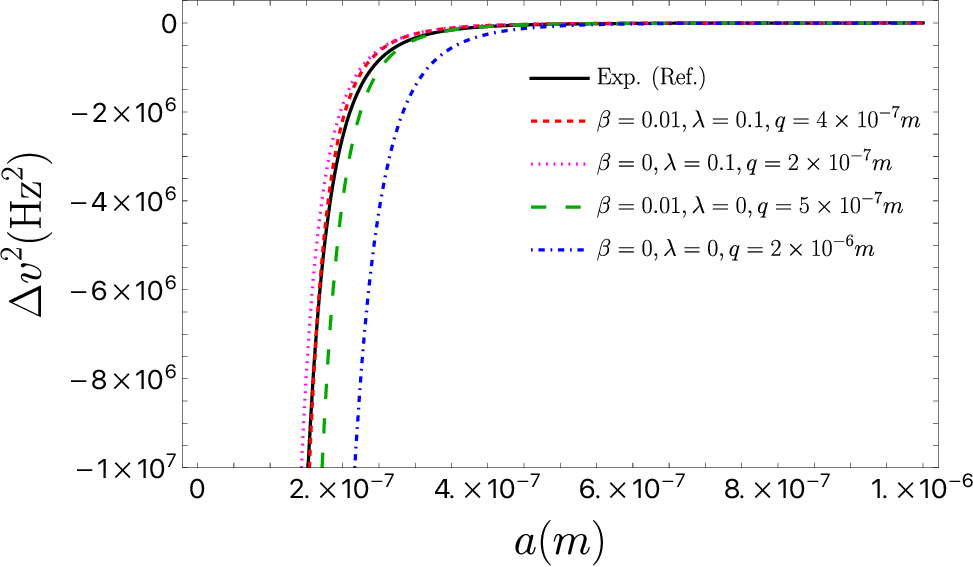}
\caption{ \label{extradim3}
Frequency shift $\Delta v^2$ with Lorentz violation in $x^3$-direction as a function of the plate's distance $a$ with parameter $\alpha = 2$. The black solid line represents experimental reference data according to Eq.\eqref{fqshiftFer}. The red dashed line corresponds to the scenario where both $\beta\neq 0$ and $\lambda\neq 0$, resulting in an estimation for the size of the extra dimension whose value is $q=4\times 10^{-7}\ m$. In contrast, the other lines correspond to cases where at least one of the parameters is set to zero: either $\beta =0$, $\lambda=0$, or both parameters are zero.}
\end{figure}

\begin{figure}[t]
\centering 
\includegraphics[width=.65\textwidth]{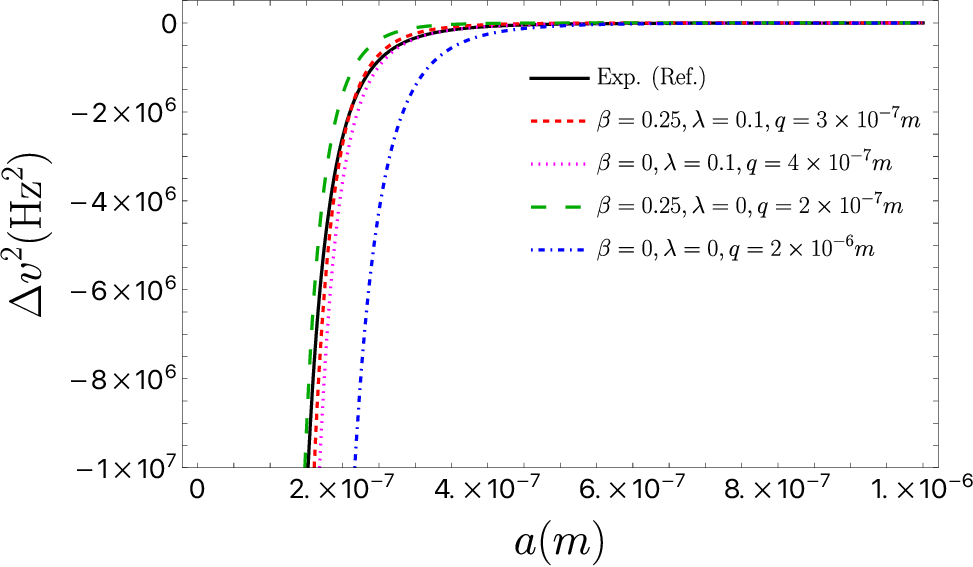}
\caption{ \label{extradim5}
Frequency shift $\Delta v^2$ with Lorentz violation in $x^5$-direction is plotted as a function of the plate's distance $a$. The black solid line is the same as in Fig.\ref{extradim3} and we have set parameter $\alpha=2$ for all lines. The red dashed line corresponds to the scenario where both $\beta\neq 0$ and $\lambda\neq 0$, resulting in an estimation for the size of the extra dimension whose value is $q=3\times 10^{-7}\ m$. In contrast, the other lines correspond to cases where at least one of the parameters is set to zero: either $\beta =0$, $\lambda=0$, or both parameters are zero. }
\end{figure}

In Figs.~\ref{extradim3} and ~\ref{extradim5}, we show the frequency shift with Lorentz violation in $x^3$- and $x^5$- directions as a function of the plate's distance $a$, respectively, where in both figures, we have set parameter $\alpha=2$, indicating that the value of $(L^2/m_{\rm eff})$ for the fermionic field is chosen to be greater than the EM field (see Eq.~\eqref{alpha}). We demonstrate the dependence of the parameters in comparison to the experimental reference data (Eq.~\eqref{fqshiftFer}).The black solid line in both figures corresponds to the experimental reference value. The red dashed line describes a scenario where both parameters $\beta \neq 0$ and $\lambda \neq 0$, yielding an estimation for the extra dimension size in $x^3$- and $x^5$- directions whose value is $q=4 \times 10^{-7}\  m$ and $q=3 \times 10^{-7}\  m$, respectively. These values agree with the range of values of cantilever experiments \cite{Murata:2014nra}. The other lines in both figures show the cases where at least one of the parameters is set to zero: either $\beta =0$, $\lambda=0$, or both parameters are zero. Note that the magenta line ($\beta=0$) in both figures could be closer to the experimental reference with a certain interval of the value for $\lambda$ and the proper value of $q$. However, as has been stated in Ref.~\cite{Pascoal:2008rna} that, this scenario with $\beta = 0$ is not suitable when determining the size of the extra dimension. On top of that, the model studied in \cite{Pascoal:2008rna} does not take into account both the quasiperiodic $\beta$ and the Lorentz-violating parameter $\lambda$ effects. We also should highlight here that we obtain a different value of the extra dimension size with one order of magnitude compared to Ref.~\cite{deFarias:2023xjf}, in which they considered the aether-EM field. 

\section{Summary and outlook}
\label{Summary}
In the present study, we have investigated the fermionic Casimir effect in the presence of compact dimensions. The topology for the system is $R^4\times S^1$, where $R^4$ corresponds to the 1+3 Minkowski spacetime and $S^1$ corresponds to extra dimensions compactified in a circle. The analysis is performed in the framework of quantum field theory with the Lorentz violation, which is parameterized by two parameters, namely, vector $u^a$ that determines the direction and $\lambda$ that determines the intensity of the violations \cite{Cruz:2018thz}. Such a model has been discussed in the context of aether field \cite{Cruz:2018thz, Carroll:2008pk, Gomes:2009ch, deFarias:2023xjf}. In the present study, we use two directions of Lorentz violation, namely, time- and space-like. For the space-like cases, we use the direction of the violation in $x^{3}$- and $x^{5}$-directions. In the system, the field in a vacuum state is confined between two parallel plates placed at the $x^{3}=0$ and $x^{3}=a$. The properties of the plates are described by MIT bag model boundary conditions \cite{Chodosetal1974a, Chodosetal1974b, Johnson1975}, which ensures that the normal current density vanishes at the boundary surfaces. For the time-like vector case and space-like vector case in $x^5$, we note that the discrete momentum $k_n(\equiv k_3 a)$ does not depend on the Lorentz violation while for the space-like vector case in $x^3$ the discrete momentum $k_n$ is affected by the Lorentz violation. It is natural because, in our system, the plates are placed at the $x^3$-axis. The compactified dimension satisfies the quasiperiodic boundary conditions, where we note that the momentum $k_5$ is discretized disrespect with Lorentz's violation and direction.  

We investigate the vacuum energy. However, we note that the vacuum energy is divergent.  To obtain the Casimir energy we use Abel-Plana like summation \cite{Romeo:2000wt}. From the obtained result, we confirm the earlier result by Ref.~\cite{Cruz:2018thz} that for the time-like vector case, the Lorentz violation does not affect the Casimir energy and pressure. In contrast, for the space-like vector case, the Lorentz violation affect both Casimir energy and pressure. We investigate the behavior of the scaled Casimir energy as well as pressure as a function of $am$, $q/a$, $\beta$, and $\lambda$, which can be described as follows. We find that both the Casimir energy and pressure goes to zero as the increasing of the parameter $am$ (see also Ref.~\cite{Cruz:2018thz}). 

We have observed that the effects of Lorentz intensity $\lambda$ in the $x^3$-direction are opposite to those in the $x^5$-direction. When considering the $x^3$-direction, the magnitude of Casimir energy decreases as the parameter $\lambda$ increases. However, in the $x^5$-direction, the magnitude of Casimir energy increases with an increase in the parameter $\lambda$. It is also worth noting that the Casimir energy is symmetric when the parameter $\beta$ changes. The maximum amplitude of the Casimir energy occurs when $\beta=0$, while the minimum amplitude occurs when $\beta=0.5$ (see Fig.~\ref{ECasalpha}). Compared to the $x^5$-direction, the Casimir energy and its pressure are more affected by Lorentz violation in the $x^3$-direction. Additionally, the minimum value of the Casimir energy magnitude in the $x^5$-direction depends on the chosen value of $\beta$. It is important to mention that when $\lambda=0$, the Casimir energy in the violation case in the $x^3$-direction gives the same value as that in the $x^5$-direction, indicating that Lorentz symmetry is preserved and has no dealings with the direction of the violation. 

 Finally, we have also investigated how to determine the size of the extra dimension utilizing the existing experimental data from the Casimir force of the EM field. For this purpose, we computed a resonant squared frequency shift $\Delta v^2$ associated with the obtained Casimir pressure for both cases. The frequency shift is proportional to $(L^{2}/m_{\text{eff}})$ whose value is determined from experimental data. In Eq.~\eqref{fqshiftFer}, we have set parameter $\alpha=2$, indicating that the value of $(L^2/m_{\rm eff})$ for the fermionic field is chosen to be greater than the EM field. 
From Figs.~\ref{extradim3} and ~\ref{extradim5}, with both parameters $\beta \neq 0$ and $\lambda \neq 0$, we found an estimation for the extra dimension size in $x^3$ and $x^5$ directions whose value is $q=4 \times 10^{-7}\  m$ and $q=3 \times 10^{-7}\  m$, respectively. 

For future work, it is interesting to apply the system for nanotubes and nanocarbon (see Ref.~\cite{Bellucci:2009hh}). In a more realistic situation, it is also interesting to investigate the thermal dependence of the Casimir energy with a similar setup (c.f. Ref.~\cite{Khoo:2011ux}).

\section*{Acknowledgments}
We thank A. Sulaksono for useful discussions and comments. A.~R. was supported by Hibah Riset PUTI Q1 Universitas Indonesia under contract No. NKB-384/UN2.RST/HKP.05.00/2024. 

\begin{appendix}


\section{Vacuum energy for the case of $x^{1}$ and $x^{2}$-directions}
\label{casex12}
In this appendix, we derive the vacuum energy for the case of the $x^1$-direction. By using the same procedure as in subsection~\ref{CasEnergySpacelike}, we can straightforwardly write down the vacuum energy as follows (cf. see Ref.~\cite{Cruz:2018thz}),
\begin{eqnarray}
E_{\rm Vac.}=-{L^2\over 2\pi^2}\int {d k_1}\int {d k_2}\sum_{\ell=-\infty}^\infty\sum^\infty_{n=1}\sqrt{m^2+(1-\lambda)^2k^2_1+k^2_2+\left({k_{n}\over a}\right)^2+k^2_{\ell}}.
\end{eqnarray}
By introducing $(1-\lambda)k_1\equiv \tilde k_1$, the above expression can be rewritten as follows
\begin{eqnarray}
E_{\rm Vac.}=-{L^2\over 2\pi^2 (1-\lambda)}\int {d \tilde k_1}\int {d k_2}\sum_{\ell=-\infty}^\infty\sum^\infty_{n=1}\sqrt{m^2+\tilde k^2_1+k^2_2+\left({k_{n}\over a}\right)^2+k^2_{\ell}}. 
\end{eqnarray}
One should note that the above expression is the form of the vacuum energy of the time-like case as in Eq.\eqref{time-like-vac} multiplied by a factor $1/(1-\lambda)$. Since we are interested in the effect of the Lorentz violations and the extra dimension on Casimir energy, we do not further consider this case. The vacuum energy of the $x^{2}$- direction can be derived with the same procedure. 
\end{appendix}

\let\doi\relax

\end{document}